\documentclass[aps,prb,twocolumn,floatfix,superscriptaddress]{revtex4-2}

\usepackage{amsmath}
\usepackage{amssymb}
\usepackage{graphicx}
\usepackage{dsfont}

\usepackage[unicode=true,
 bookmarks=true,bookmarksnumbered=false,bookmarksopen=false,
 breaklinks=false,pdfborder={0 0 1},backref=false,colorlinks=true]
 {hyperref}
\hypersetup{
 linkcolor=magenta, urlcolor=blue, citecolor=blue, pdfstartview={FitH}, unicode=true}

\usepackage{amsfonts}\usepackage{tabularx}\usepackage{dcolumn}\usepackage{bm}\usepackage{graphicx}
\usepackage{epstopdf}
\usepackage{xcolor}
\hypersetup{urlcolor=blue}
\usepackage{times}
\usepackage{hyperref}

\usepackage{float}
\makeatletter
\let\newfloat\newfloat@ltx
\makeatother

\usepackage{algorithm}
\usepackage{algorithmic}

\usepackage{placeins}

\begin{document}

\title{Transformer Quantum State: A Multi-Purpose Model for Quantum Many-Body Problems}
\author{Yuan-Hang Zhang}
\email{yuz092@ucsd.edu}
\affiliation{Department of Physics, University of California, San Diego, CA 92093, USA}
\author{Massimiliano Di Ventra}
\email{diventra@physics.ucsd.edu}
\affiliation{Department of Physics, University of California, San Diego, CA 92093, USA}

\begin{abstract}
Inspired by the advancements in large language models based on transformers, we introduce the transformer quantum state (TQS): a versatile machine learning model for quantum many-body problems. In sharp contrast to Hamiltonian/task specific models, TQS can generate the entire phase diagram, predict field strengths with experimental measurements, and transfer such a knowledge to new systems it has never been trained on before, all within a single model. With specific tasks, fine-tuning the TQS produces accurate results with small computational cost. Versatile by design, TQS can be easily adapted to new tasks, thereby pointing towards a general-purpose model for various challenging quantum problems.

\end{abstract}
\maketitle

\section{Introduction}

Determining the state of a quantum many-body system is one of the fundamental problems in physics. While the exponential growth of the Hilbert space precludes brute-force calculations, computational methods such as quantum Monte Carlo \cite{foulkes2001quantum} and tensor network-based methods \cite{schollwock2011density} allow for efficient simulations of certain problems, each with their own strengths and weaknesses. 

More recently, the advancements in machine learning techniques and models have influenced the physics community. In fact, the introduction of neural networks (NNs) as variational states for quantum many-body problems has greatly expanded the types and sizes of systems that can be efficiently tackled. For instance, the restricted Boltzmann machine \cite{hopfield1982neural, hinton2002training} was the first NN model applied to correlated quantum systems \cite{carleo2017solving}, followed by models with different architectures such as feed-forward \cite{cai2018approximating, choo2018symmetries}, convolutional \cite{liang2018solving, choo2019two}, recurrent \cite{hibat2020recurrent}, and autoregressive \cite{sharir2020deep, barrett2022autoregressive, luo2021gauge} ones. With the ability to encode area and volume-law entanglement \cite{deng2017quantum}, NNs are especially advantageous in dealing with high-dimensional systems. And with proper tricks, they can also greatly ease the fermion sign problem \cite{cai2018approximating}. Yet, despite these successes, the previous approaches are limited to specific tasks.

Recently, a new task-agnostic model has been put forward by the machine learning community: the transformer architecture \cite{vaswani2017attention}. Since its introduction, this model has dominated the field by achieving state-of-the-art results in almost every natural language processing task \cite{devlin2018bert, radford2018improving, radford2019language, brown2020language}, thus rendering the recurrent neural networks obsolete in merely a few years. Transformers have also been adapted to different tasks such as image recognition \cite{dosovitskiy2020image}, audio processing \cite{dong2018speech} and graph classification \cite{velickovic2018graph}, all achieving remarkable results. 

This feat relies on an impressive aspect of transformer models: their ability to scale to very large sizes~\cite{brown2020language,reed2022generalist}. 
%As a notable example, the language model GPT-3 \cite{brown2020language}, has 175 billion parameters and is trained on 45 TB of text data. A much larger model size implies capability to handle a much wider variety of data, which encourages knowledge transfer between different tasks. For GPT-3, while the training goal is nothing more than predicting the next token in a sequence, it is capable of a wide range of tasks, including writing articles, solving math problems and generating computer codes. More recently, the generalist agent Gato \cite{reed2022generalist} could play Atari games, control robot arms, caption images and chat, all within the same neural sequence model. 
When facing with a new task, few-shot learning \cite{brown2020language} allows a general purpose model to easily adapt with merely a few examples in natural language. And when better performance is desired, fine-tuning on a small dataset produces satisfactory results within a short time \cite{radford2019language}. 

These results give hope that such an architecture may be of great help in quantum physics as well. However, the application of transformers in this field is still rather limited, with a few results concerning quantum lattice models \cite{luo2021gauge}, open systems \cite{luo2022autoregressive}, quantum state tomography \cite{cha2021attention} and quantum circuit simulation \cite{carrasquilla2021probabilistic}, while the task-agnostic property is barely used. Therefore, the full potential of the transformer architecture has yet to be explored. 

Contrary to the general-purpose models mentioned above, NN models in physics are usually highly specialized, serving a single purpose such as representing wave functions \cite{carleo2017solving, cai2018approximating, choo2018symmetries, liang2018solving, choo2019two, hibat2020recurrent, sharir2020deep, barrett2022autoregressive}, preparing and controlling quantum states \cite{carrasquilla2021probabilistic, zhang2020topological}, recognizing phase transitions \cite{carrasquilla2017machine, van2017learning}, realizing quantum state tomography \cite{torlai2018neural, cha2021attention, zhang2021efficient}, etc. Such tasks share a lot of common knowledge, making it ideal to have a single, unified model that handles them all, with the possibility of discovering new physics at the intersection of different tasks. 

\begin{figure*}[htbp]
	\centering
	\includegraphics[width = 0.97\textwidth]{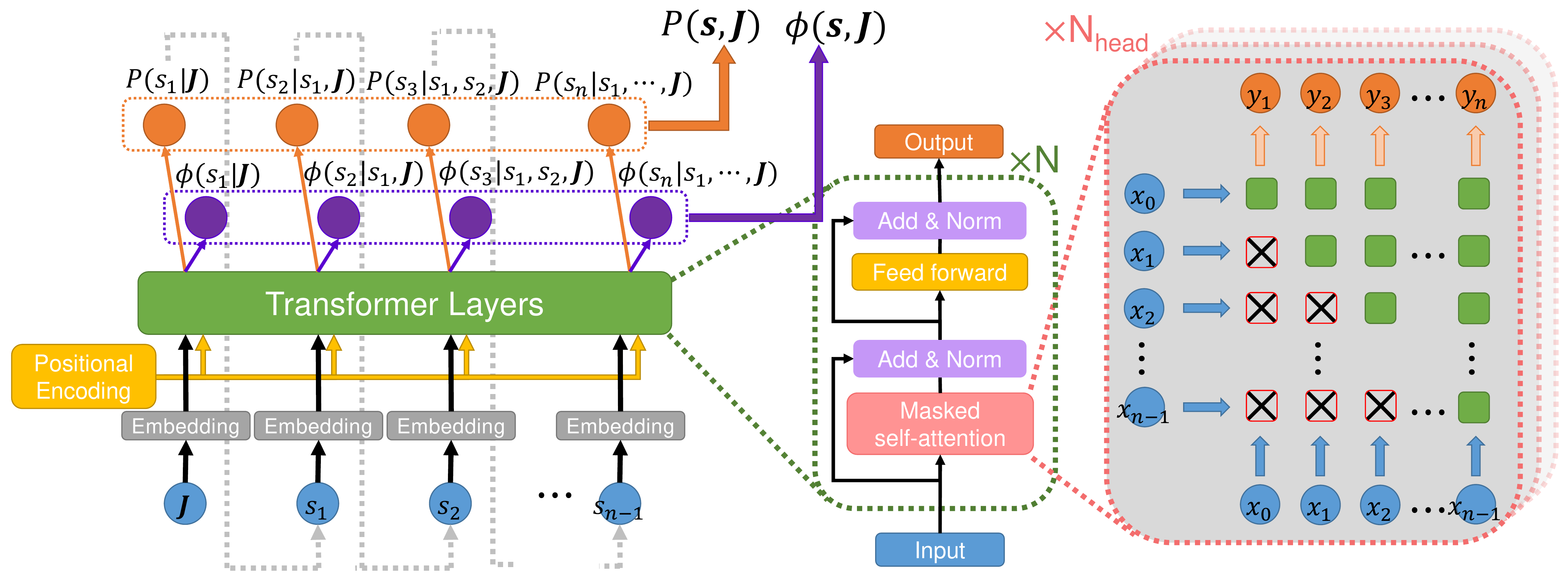}
	\caption{The structure of a TQS. Left: the overall architecture of our model. We use the standard encoder-only transformer architecture, utilizing an embedding layer to map different inputs into a single unified feature space, and pass them through $N$ identical transformer encoder blocks, followed by two different output heads, parameterizing the amplitude $P$ and phase $\phi$, respectively. Middle: The structure of a transformer encoder block. Right: the mask structure in a masked self-attention operator. Squares with a cross represent the masks, blocking the flow of information, so that each site only has access to its predecessors. This ensures that the autoregressive property is satisfied. }
	\label{fig:transformer}
\end{figure*}

As a first step, we may consider using NNs as variational wave functions. Traditionally, each NN can only represent a specific quantum state, and tasks such as generating a phase diagram requires retraining of the same NN from scratch for hundreds of times, even if nearby data points have similar features. 

%A possible improvement on this issue is transfer learning, namely reusing the features learned on one task, applying them to similar tasks to accelerate training. For instance, Ref.~\cite{zen2020transfer} demonstrated the effectiveness of this approach on restricted Boltzmann machines. 

In this paper, we consider a different perspective: instead of modeling a specific quantum state, we attempt to represent a {\it family of quantum states} within a single neural network. More precisely, we focus on the joint distribution of the wave function and relevant physical parameters such as interaction strength, external field, and/or system size. For the underlying NN, we choose the transformer architecture for its versatility and strong performance across different tasks. 

We call this model a {\it transformer quantum state} (TQS), and show that it is capable of generating the entire phase diagram of a many-body system, predicting field strengths with as few as one experimental measurement, and transferring knowledge to new systems it has never seen before, all within a single model.

\section{Transformer Quantum State}
\label{sec:tqs}
% \emph{Transformer quantum state} \textemdash 
Consider the probability distribution $P(\mathbf{s}, \mathbf{J})\equiv P(s_1, \cdots, s_n, J_1, \cdots, J_m)$, where $s_i\in\{0, 1, \cdots, d-1\}$ are discrete variables representing the physical degrees of freedom such as spin or occupation number, and $J_j$ correspond to other physical parameters, either continuous or discrete. Such a state space grows exponentially with the number of variables, and a compact representation is desired. 

To represent $P(\mathbf{s}, \mathbf{J})$, we adopt the transformer architecture, and autoregressively model the entire distribution as a product of conditional distributions, 
\begin{equation}
    P(\mathbf{s}, \mathbf{J}) = P(\mathbf{J})\prod_{i=1}^n P(s_i|s_1,\cdots, s_{i-1}, \mathbf{J}). \label{eq:autoregressive}
\end{equation}

The structure of the transformer is shown in Fig.~\ref{fig:transformer}, with each output of the neural network representing one of the conditional distributions. For a detailed explanation of the transformer architecture, see Appendix~\ref{sec:implementation}. 

Contrary to energy-based models such as restricted Boltzmann machines \cite{carleo2017solving}, the autoregressive structure allows for efficient sampling \cite{sharir2020deep}. Since each conditional probability $P(s_i|s_1,\cdots, s_{i-1}, \mathbf{J})$ does not depend on any variable $s_j$ with $j>i$, starting from $s_1$, one can sequentially sample $s_i$ according to the previously sampled configurations, using the $i$-th conditional distribution only. Using the idea developed in \cite{barrett2022autoregressive}, efficiency of the sampling algorithm can be further improved by only sampling unique configuration strings, and the details are explained in Appendix~\ref{sec:sampling}. 

We assume that $\mathbf{J}$ has a predefined prior distribution $P(\mathbf{J})$, which, in general, can be chosen as a uniform distribution over the range of interest (e.g., to study the transition in the Heisenberg $J_1$-$J_2$ model \cite{schulz1996magnetic, choo2019two}, one can fix $J_1=1$ and make $J_2$ uniform over $[0, 1]$). 

Our aim is to model quantum states $|\psi(\mathbf{J})\rangle$, which are complex-valued quasiprobability distributions. To this end, we expand them in the computational basis and separate their amplitude $A$ and phase $\phi$,
\begin{equation}
\begin{aligned}
    |\psi(\mathbf{J})\rangle &= \sum_{\mathbf{s}} \psi(\mathbf{s}, \mathbf{J})|\mathbf{s}\rangle \\
    &= \sum_{\mathbf{s}} A(\mathbf{s}, \mathbf{J}) \exp(i\phi(\mathbf{s}, \mathbf{J}))|\mathbf{s}\rangle.
\end{aligned}
\end{equation}

Since squared amplitude has the probability interpretation, we choose 
\begin{equation}
    A(\mathbf{s}, \mathbf{J}) = \sqrt{P(\mathbf{s}, \mathbf{J})},
\end{equation}
with $P(\mathbf{s}, \mathbf{J})$ specified in Eq.~\eqref{eq:autoregressive}. The phase $\phi$ has no restrictions and can be either positive or negative, and we represent it with a similar autoregressive structure:
\begin{equation}
    \phi(\mathbf{s}, \mathbf{J}) = \sum_i \phi(s_i|s_1,\cdots, s_{i-1}, \mathbf{J}).
\end{equation}

\subsection{Ground state of a family of Hamiltonians}

% \emph{Ground state of a family of Hamiltonians} \textemdash 
The first task we consider is finding the ground state $|\psi\rangle$ of many-body Hamiltonians. Per the standard procedure, this can be done by minimizing the variational energy estimation, $\langle \psi | \hat{H} |\psi\rangle$, over the target Hamiltonian $\hat{H}$. A minor complication is that, instead of a single Hamiltonian $\hat{H}$, we have now a {\it family} of Hamiltonians $\{\hat{H}(\mathbf{J})\}$. In Appendix~\ref{sec:optimization}, we show that the family of ground states $|\psi(\mathbf{J})\rangle$ corresponds to the ground state $|\Psi\rangle$ of the super-Hamiltonian $\hat{\mathcal{H}}=\bigoplus_{\mathbf{J}}\frac{\hat{H}(\mathbf{J})}{|E_g(\mathbf{J})|}$ in the extended Hilbert space, and we can optimize the TQS by minimizing $\langle \Psi |\hat{\mathcal{H}} |\Psi\rangle$, which follows the standard procedure.

Once we have the family of ground states $\psi(\mathbf{s}, \mathbf{J})$, an immediate application is to estimate the physical parameters $\mathbf{J}$ using samples $\mathbf{s}$ from the wave function. This follows trivially from the conditional probability: 

\begin{equation}
    P(\mathbf{J}|\mathbf{s}) = \frac{P(\mathbf{s}, \mathbf{J})}{P(\mathbf{s})}.
\end{equation}

In practice, given a set of measurements $\{\mathbf{s}_k\}$, $\mathbf{J}$ can be predicted using standard maximum likelihood estimation \cite{myung2003tutorial}, by maximizing the log-likelihood functional,

\begin{equation}
    \mathcal{L}(\mathbf{J}) = \sum_k \log P(\mathbf{s}_k| \mathbf{J}). \label{eq:log_likelihood}
\end{equation}
In this way, we can efficiently determine physical properties of a quantum system with few measurements. Details of the implementation can be found in Appendix~\ref{sec:param}.

This task is somewhat similar to shadow tomography \cite{aaronson2019shadow, huang2020predicting}, in the sense that we are predicting properties of a quantum system with a few measurements, but with more restrictions and with certain prior knowledge required. On the other hand, Ref.~\cite{carrasquilla2017machine} considered another similar task of recognizing phases from measurements using machine learning, which is formulated as a classification task. In comparison, our task falls in the middle of the two mentioned above, and to the best of our knowledge, it has never been proposed. Under this setting, the TQS can handle this task extremely efficiently. In fact, with the prior knowledge that a quantum state $|\psi\rangle$ comes from a family of states $|\psi(\mathbf{J})\rangle$, we can efficiently determine the physical parameters $\mathbf{J}$, with as few as one measurement only. 

Furthermore, we show that the TQS can transfer knowledge to new systems it has never seen before. This follows the pre-training plus fine-tuning methodology commonly adopted in natural language models \cite{devlin2018bert, radford2018improving}. In the zero-shot setting \cite{brown2020language}, after training on the family of Hamiltonians $\hat{H}(\mathbf{J})$, TQS can generate the ground state of new Hamiltonians $\hat{H}(\mathbf{J^*})$ with $\mathbf{J^*} \notin \{\mathbf{J}\}$, albeit with slightly larger error. When higher accuracy is desired, one can fine-tune the TQS on the specific Hamiltonian $\hat{H}(\mathbf{J}^*)$, to obtain accurate results within a much shorter time comparing to learning from scratch. 

\section{Results}
% \emph{Results} \textemdash 
As a prototypical test bed, we first examine the 1D transverse field Ising (TFI) model, whose Hamiltonian is
\begin{equation}
    \hat{H} = -J \sum_{i=1}^{n-1}\sigma^z_i \sigma^z_{i+1} - h\sum_{i=1}^n\sigma^x_i, \label{eq:Ising}
\end{equation}
where $J$ is the coupling constant and $\sigma^z$ and $\sigma^x$ are Pauli matrices. 
In Appendix~\ref{sec:XYZ}, we also provide numerical results on the 1D XYZ model. 

\subsection{Ground state calculations}

To begin with, we pre-train the TQS on the family of TFI Hamiltonians $\hat{H}(n, h)$, specified in Eq.~\eqref{eq:Ising}. We fix $J=1$, and assume a uniform distribution of the transverse field $h\in[0.5, 1.5]$. The system size $n$ can take any even integer value with equal probability in the range of $[10, 40]$. We explicitly enforced parity and spin flip symmetry on the TQS, with details elaborated in Appendix~\ref{sec:symmetry}. 

\begin{figure}[thbp]
	\centering
	\includegraphics[width = 0.45\textwidth]{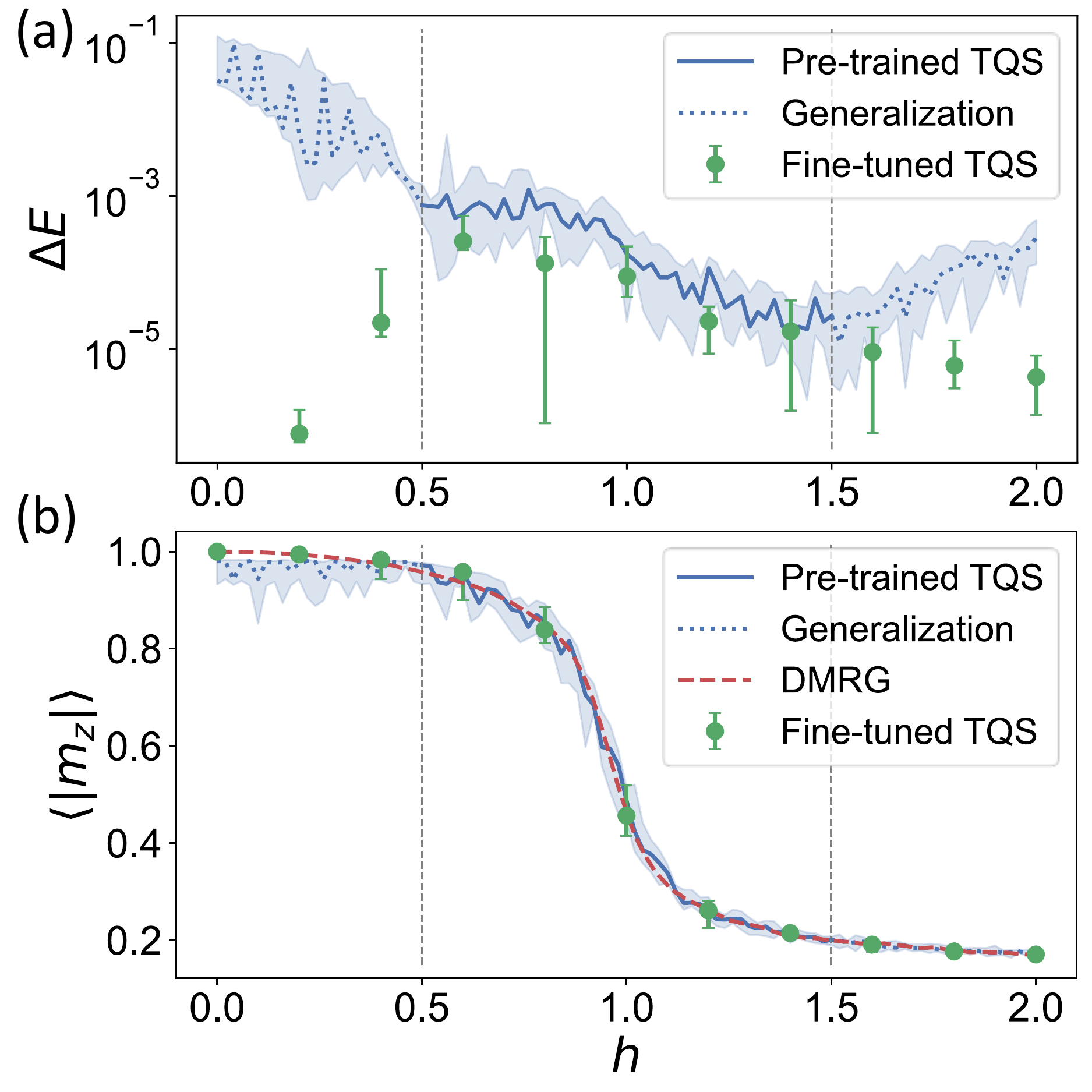}
	\caption{Results on the ground state of the TFI Hamiltonian, Eq.~\eqref{eq:Ising}, with $n=40$. Lines and data points are medians of 10 estimations, while shaded regions and error bars enclose $10^{\mathrm{th}}$ to $90^{\mathrm{th}}$ percentile. Dotted lines are generalizations to regions TQS has not been trained on. (a) The relative error of the ground state energy, $\Delta E = |(E-E_{\mathrm{ground}}) / E_{\mathrm{ground}}|$. $E_{\mathrm{ground}}$ is estimated with DMRG, which is accurate up to $10^{-10}$. (b) Absolute value of the magnetization along the $z$ direction, $\langle|m_z|\rangle$. We can observe the transition near $h=1$. } 
	\label{fig:fields}
\end{figure}

After pre-training for $10^5$ iterations, we plot the ground state energy, $E$, and magnetization along the $z$ direction, $m_z=\sum_i\langle\sigma^z_i\rangle / N$, for $n=40$, $h\in[0, 2]$, in Fig.~\ref{fig:fields}. Since we explicitly symmetrized the TQS with the $|0\rangle\leftrightarrow|1\rangle$ spin flip symmetry, we always have $\langle m_z\rangle=0$, so $\langle|m_z|\rangle$ is plotted instead.  Note that while the TQS is only trained in the range of $h\in[0.5, 1.5]$, it can infer the properties of the ground state when $h\in[0, 0.5)$ and $h\in(1.5, 2]$ with slightly larger error, without any additional inputs except the value of $h$. 

\begin{figure}[htbp]
	\centering
	\includegraphics[width = 0.45\textwidth]{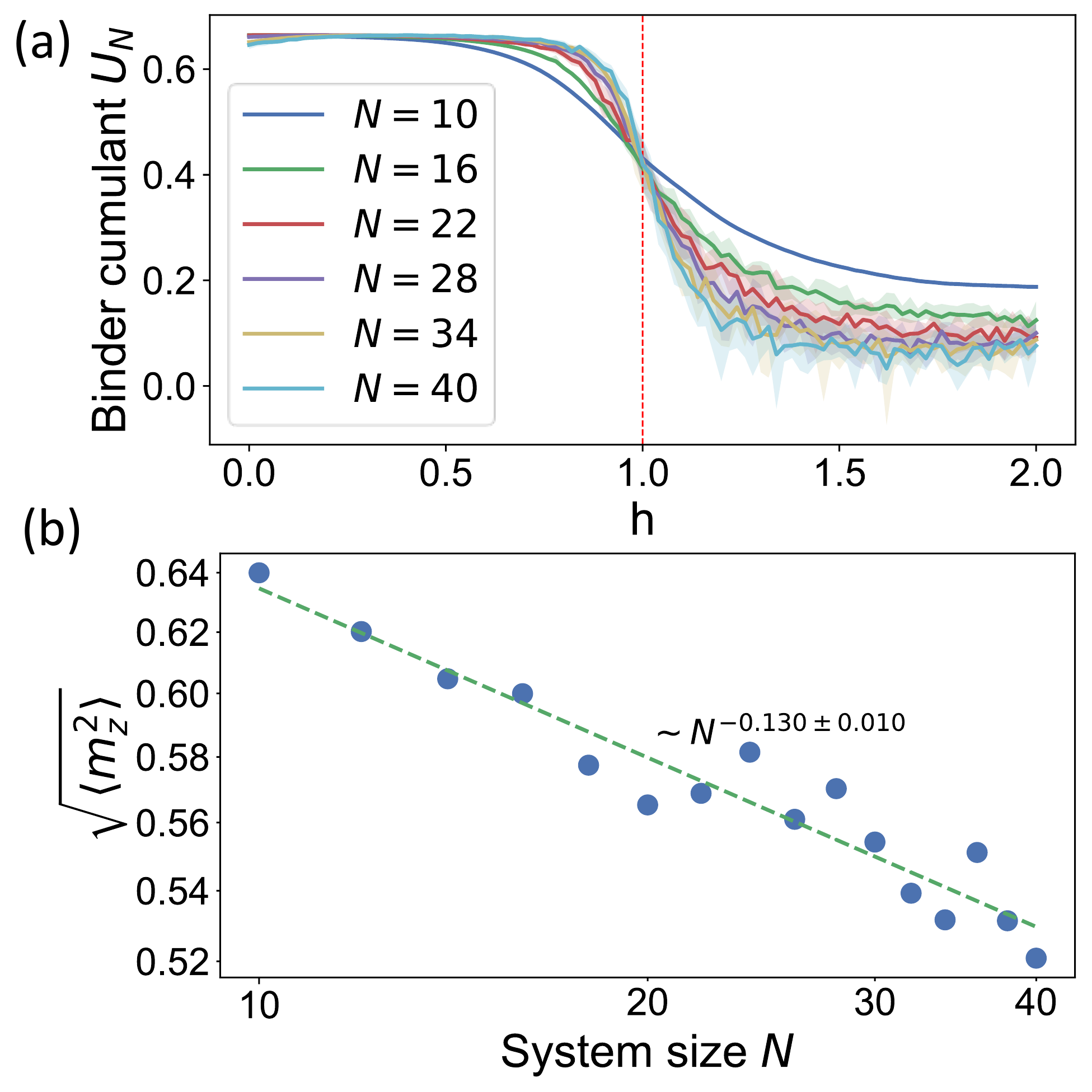}
	\caption{Finite-size scaling calculations on the TFI model, using the TQS trained with $h\in[0.5, 1.5]$. (a) Binder cumulant \cite{binder1981finite}, $U_N=1-\frac{\langle m_z^4\rangle_N}{3\langle m_z^2\rangle_N^2}$, plotted for various system sizes $N$. At the critical point $h_c$, $U_N$ is invariant with the system size $N$, and finding the crossing of various $U_N$ curves can help us determine the critical point. In this figure, we identify $h_c=1$, which agrees with the theoretical prediction. (b) Finite-size scaling of the mean-square-root magnetization \cite{pang2019critical} at the critical point $h=1$. Using the finite-size scaling ansatz \cite{suzuki2012quantum}, at the critical point, $\sqrt{\langle m_z^2\rangle}|_{h_c}\sim N^{-\beta/\nu}$. A linear fit on the log-log scale gives $\beta/\nu=0.130\pm0.010$, which matches the theoretical values $\beta=1/8$ and $\nu=1$. } 
	\label{fig:finite_size}
\end{figure}

Finite-size scaling can be easily carried out using TQS. With a variable input length, we can represent an arbitrary number of degrees of freedom within a single TQS model. Using the same model trained with $h\in[0.5, 1.5]$, in Fig.~\ref{fig:finite_size} we show that finite-size scaling analysis on the TFI model correctly identifies the phase transition point $h=1$, and the predicted critical exponents satisfy $\beta / \nu = 0.130\pm 0.010$, which match the theoretical predictions $\beta=1/8$, $\nu=1$. Details of the calculation can be found in Appendix~\ref{sec:finite_size}. 

\begin{figure}[thbp]
	\centering
	\includegraphics[width = 0.45\textwidth]{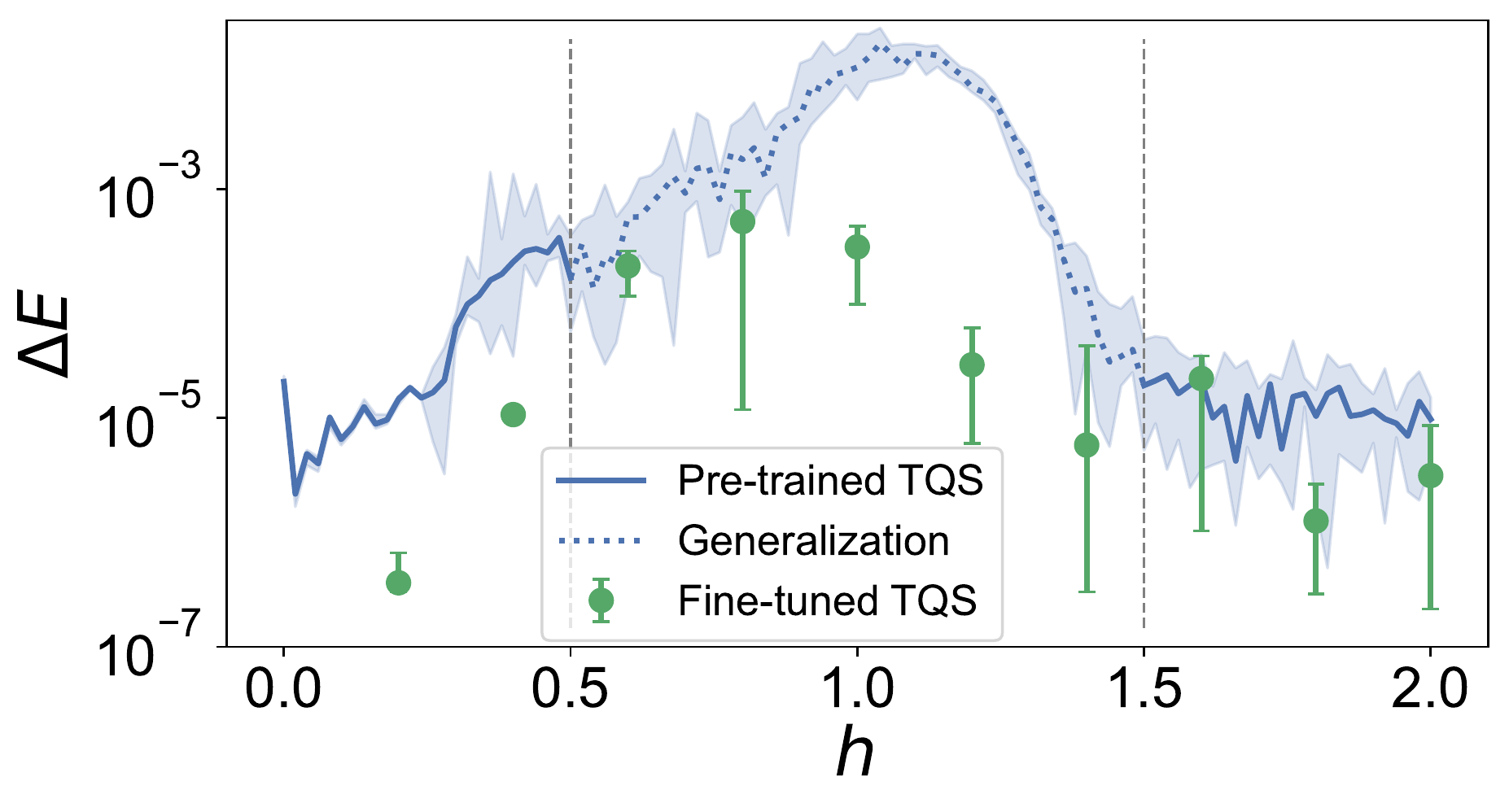}
	\caption{Relative error of the ground state energy of the TFI Hamiltonian, Eq.~\eqref{eq:Ising}, with $n=40$. Lines and data points are medians of 10 estimations, while shaded regions and error bars enclose $10^{\mathrm{th}}$ to $90^{\mathrm{th}}$ percentile. Dotted lines are generalizations to regions TQS has not been trained on. Data points below $10^{-7}$ are not shown for a clearer illustration. }
	\label{fig:phase_transition}
\end{figure}

Similar experiments are carried out where the TQS is trained in the range $h\in[0, 0.5]\cup[1.5, 2]$, and the results are shown in Fig.~\ref{fig:phase_transition}. Although training is only carried out either deep in the ferromagnetic phase or paramagnetic phase, TQS can still infer the ground state energy and magnetization of TFI near the phase transition with reasonable accuracy.

However, in Appendix~\ref{sec:finite_size} we show that, this interpolated state undergoes phase transition at $h=1.24$ instead of $h=1$, with critical exponents different from the usual Ising transition. Without access to training data near the phase boundary, TQS cannot accurately predict the phase transition. Rather, it generates a fictitious physical system with its own critical behaviors.

At this point, we further fine-tune the TQS on specific points $\hat{H}(n^*, h^*)$ for an additional $2\times 10^3$ iterations, and the results are also shown in Figs.~\ref{fig:fields}, \ref{fig:phase_transition}. Outside of the pre-trained region, the accuracy improved dramatically up to a few orders of magnitudes. Within the pre-trained region, there is also a small improvement in accuracy, but not as much since the pre-trained model already works well.
% The energy estimation did not improve much within the pre-trained region $h\in[0.5, 1.5]$, which is likely due to the fact that the pre-trained model already performs well, and that optimization gets more difficult near the point of phase transition $h=1$. 

\begin{figure}[b]
	\centering
	\includegraphics[width = 0.45\textwidth]{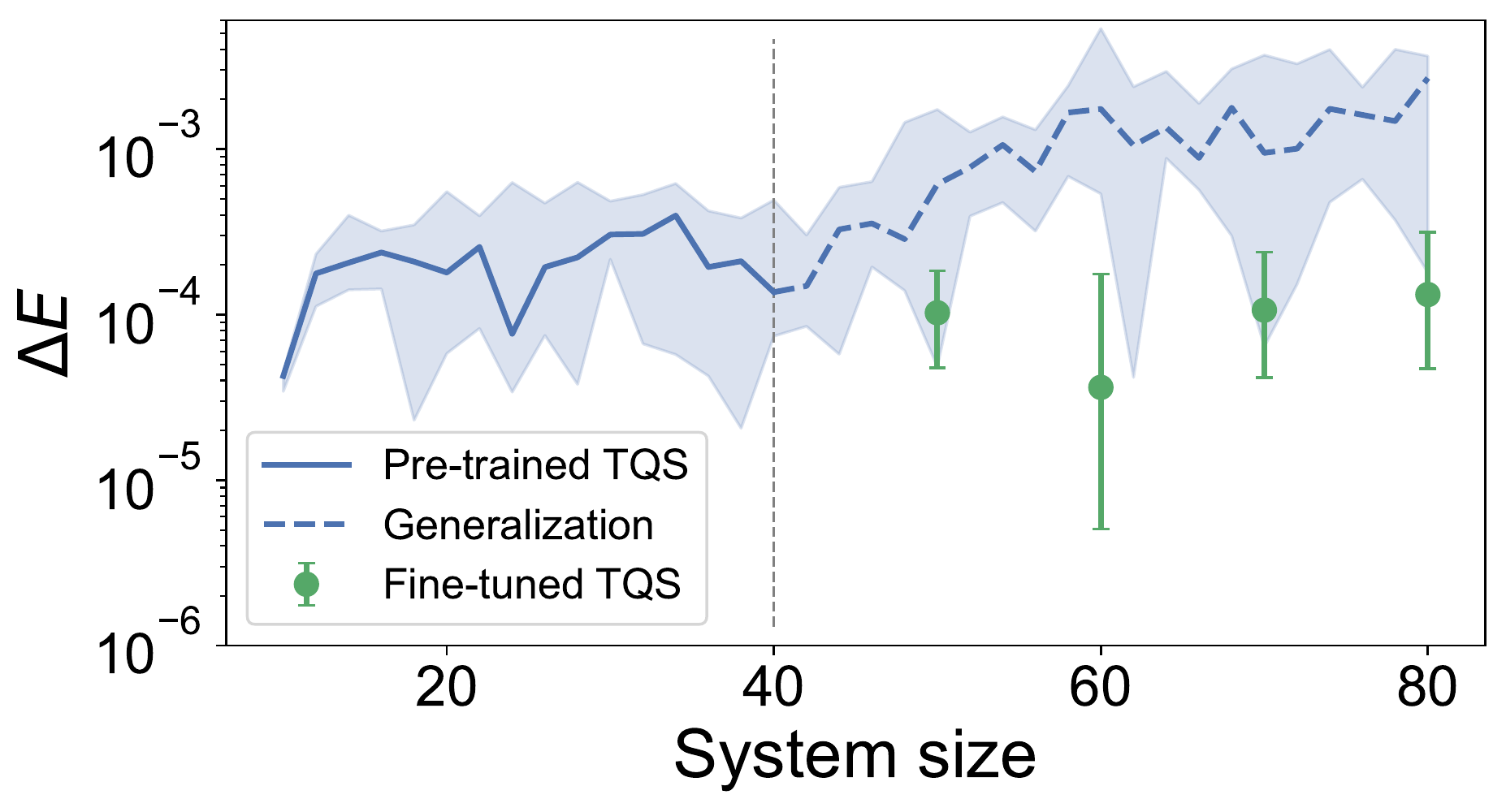}
	\caption{Relative error of the ground state energy of the TFI Hamiltonian, Eq.~\eqref{eq:Ising}, with $h=1$. Lines and data points are medians of 10 estimations, while shaded regions and error bars enclose $10^{\mathrm{th}}$ to $90^{\mathrm{th}}$ percentile. Dashed lines are generalizations to regions TQS has not been trained on. The pre-trained TQS can infer the ground state energy of much larger systems than what it is trained on, without any additional input except the system size $n$, albeit with slightly larger error. By fine-tuning with an additional $2\times 10^3$ iterations, the accuracy improves by an order of magnitude. }
	\label{fig:sizes}
\end{figure}

As a further test, we fix $h=1$, and compute the ground state energy of systems with different sizes $n\in[10, 80]$ (using the model trained in $h\in[0.5, 1.5]$). The result is plotted in Fig.~\ref{fig:sizes}. Again, even if the pre-trained model has never seen any system with more than 40 spins, it can generalize to much larger systems, and their energy estimations can be greatly improved by fine-tuning for an additional $2\times 10^3$ iterations.

\subsection{Predicting parameters}

Next, with the learned distribution $P(\mathbf{s}, n, h)$, we want to predict the transverse field $h$ using experimentally available measurements. To this end, we simulate the experiment by computing the ground state of the TFI model using the density matrix renormalization group (DMRG) \cite{schollwock2011density, tenpy}, and generate a synthetic dataset with projective measurements in the computational basis. Details of DMRG calculations can be found in Appendix~\ref{sec:DMRG}.

\begin{figure}[t]
	\centering
	\includegraphics[width = 0.45\textwidth]{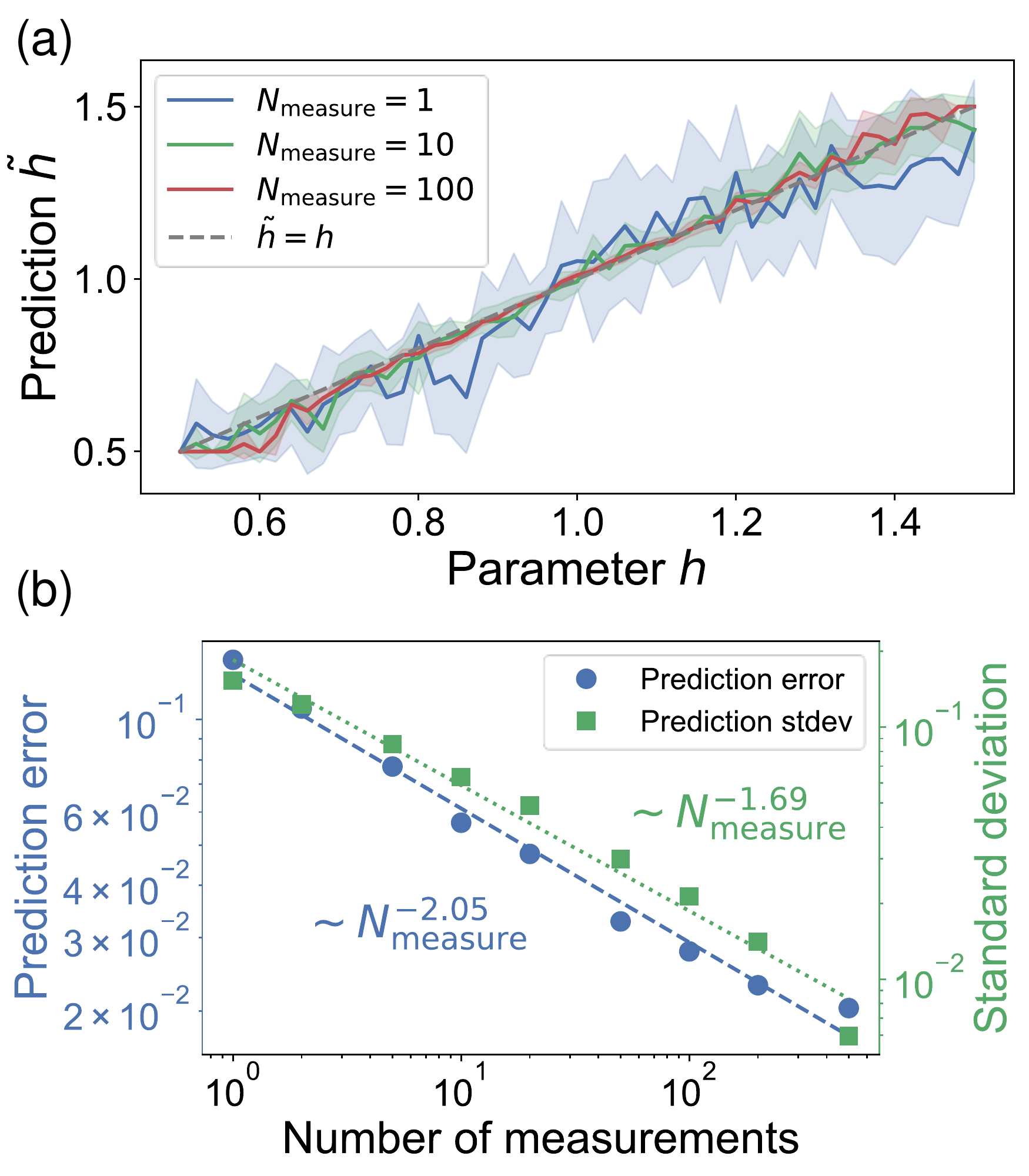}
	\caption{(a) The predicted field strength $\tilde{h}$ vs. the actual field strength $h$, with varying number of measurements. Solid lines are mean values of 10 predictions, and shaded regions enclose one standard deviation. The dashed line represents the expected result, $\tilde{h}=h$. (b) Scaling of the prediction error, $|\tilde{h}-h|$, and standard deviation, $\sigma_{\tilde{h}}$, vs. the number of measurements. Each data point is computed with 10 predictions. We observe an empirical power law scaling, with $|\tilde{h}-h|\sim N_{\mathrm{measure}}^{-2.05}$ and $\sigma_{\tilde{h}}\sim N_{\mathrm{measure}}^{-1.69}$. }
	\label{fig:predictions}
\end{figure}

We fix $n=40$, and predict $h$ by maximizing Eq.~\eqref{eq:log_likelihood}, the log-likelihood functional, with varying number of measurements. The results are shown in Fig.~\ref{fig:predictions}. Surprisingly, with as few as one measurement, TQS gives reasonable estimations of $h$. Increasing the number of measurements improves the quality of prediction, and an empirical power law scaling of the prediction error versus the number of measurements is observed. 

\section{Discussion}
% \emph{Discussion} \textemdash 
In summary, our results demonstrate how the TQS learns various ground state properties of a physical system, and appropriately uses the acquired knowledge to solve new problems. 
TQS marks the first step towards a general purpose model for quantum physics. Although we only explored here the ground states of many-body Hamiltonians, it is possible to encode many additional operations and information into the TQS, such as unitary transformations, time evolution, positive operator-valued measurements, etc. Thanks to the flexibility of neural sequence models and the transformer architecture, all the additional information can be formulated as new tokens to be passed into the embedding layer, thus maintaining the model structure simple and unified. 

%Moreover, our formulation of TQS is compatible with state-of-the-art language models \cite{brown2020language, radford2019language, radford2018improving}, and it is possible to fine-tune a physics model on top of a language model such as GPT-3 \cite{brown2020language}. This would make it possible for models to ``comprehend'' physical tasks using natural language, enabling prompt-based problem solving \cite{brown2020language}, and avoiding the inconsistencies in hand-designed features and formats for various different tasks, thus leading towards a unified general purpose model. 

Limited by available computational resources, we were unable to train larger models for a wider range of tasks. But we believe that, with the advancements in the development of new computing paradigms such as MemComputing \cite{di2022memcomputing}, such models can be pushed even further. This would help researchers understand various challenging quantum phenomena, and assist them in the design and characterization of near-term quantum devices. 
%Moreover, combined with the latest advancements in problem-solving and commonsense reasoning with large language models, it won't be a surprise to see the birth of an AI scientist in the near future. 

\emph{Data availability---}The code for all simulations performed in this paper, the weights of a pre-trained TQS on the Ising model, and a synthetic dataset generated using DMRG are available at \href{https://github.com/yuanhangzhang98/transformer_quantum_state}{https://github.com/yuanhangzhang98/transformer\_quantum\_state}.

\emph{Acknowledgements}\textemdash
We acknowledge financial support from the Department of Energy under Grant No. DE-SC0020892.

\bibliographystyle{apsrev4-2}
\bibliography{references}

\begin{thebibliography}{53}%
\makeatletter
\providecommand \@ifxundefined [1]{%
 \@ifx{#1\undefined}
}%
\providecommand \@ifnum [1]{%
 \ifnum #1\expandafter \@firstoftwo
 \else \expandafter \@secondoftwo
 \fi
}%
\providecommand \@ifx [1]{%
 \ifx #1\expandafter \@firstoftwo
 \else \expandafter \@secondoftwo
 \fi
}%
\providecommand \natexlab [1]{#1}%
\providecommand \enquote  [1]{``#1''}%
\providecommand \bibnamefont  [1]{#1}%
\providecommand \bibfnamefont [1]{#1}%
\providecommand \citenamefont [1]{#1}%
\providecommand \href@noop [0]{\@secondoftwo}%
\providecommand \href [0]{\begingroup \@sanitize@url \@href}%
\providecommand \@href[1]{\@@startlink{#1}\@@href}%
\providecommand \@@href[1]{\endgroup#1\@@endlink}%
\providecommand \@sanitize@url [0]{\catcode `\\12\catcode `\$12\catcode
  `\&12\catcode `\#12\catcode `\^12\catcode `\_12\catcode `\%12\relax}%
\providecommand \@@startlink[1]{}%
\providecommand \@@endlink[0]{}%
\providecommand \url  [0]{\begingroup\@sanitize@url \@url }%
\providecommand \@url [1]{\endgroup\@href {#1}{\urlprefix }}%
\providecommand \urlprefix  [0]{URL }%
\providecommand \Eprint [0]{\href }%
\providecommand \doibase [0]{https://doi.org/}%
\providecommand \selectlanguage [0]{\@gobble}%
\providecommand \bibinfo  [0]{\@secondoftwo}%
\providecommand \bibfield  [0]{\@secondoftwo}%
\providecommand \translation [1]{[#1]}%
\providecommand \BibitemOpen [0]{}%
\providecommand \bibitemStop [0]{}%
\providecommand \bibitemNoStop [0]{.\EOS\space}%
\providecommand \EOS [0]{\spacefactor3000\relax}%
\providecommand \BibitemShut  [1]{\csname bibitem#1\endcsname}%
\let\auto@bib@innerbib\@empty
%</preamble>
\bibitem [{\citenamefont {Foulkes}\ \emph {et~al.}(2001)\citenamefont
  {Foulkes}, \citenamefont {Mitas}, \citenamefont {Needs},\ and\ \citenamefont
  {Rajagopal}}]{foulkes2001quantum}%
  \BibitemOpen
  \bibfield  {author} {\bibinfo {author} {\bibfnamefont {W.}~\bibnamefont
  {Foulkes}}, \bibinfo {author} {\bibfnamefont {L.}~\bibnamefont {Mitas}},
  \bibinfo {author} {\bibfnamefont {R.}~\bibnamefont {Needs}},\ and\ \bibinfo
  {author} {\bibfnamefont {G.}~\bibnamefont {Rajagopal}},\ }\href
  {https://journals.aps.org/rmp/abstract/10.1103/RevModPhys.73.33} {\bibfield
  {journal} {\bibinfo  {journal} {Reviews of Modern Physics}\ }\textbf
  {\bibinfo {volume} {73}},\ \bibinfo {pages} {33} (\bibinfo {year}
  {2001})}\BibitemShut {NoStop}%
\bibitem [{\citenamefont {Schollw{\"o}ck}(2011)}]{schollwock2011density}%
  \BibitemOpen
  \bibfield  {author} {\bibinfo {author} {\bibfnamefont {U.}~\bibnamefont
  {Schollw{\"o}ck}},\ }\href
  {https://www.sciencedirect.com/science/article/pii/S0003491610001752}
  {\bibfield  {journal} {\bibinfo  {journal} {Annals of physics}\ }\textbf
  {\bibinfo {volume} {326}},\ \bibinfo {pages} {96} (\bibinfo {year}
  {2011})}\BibitemShut {NoStop}%
\bibitem [{\citenamefont {Hopfield}(1982)}]{hopfield1982neural}%
  \BibitemOpen
  \bibfield  {author} {\bibinfo {author} {\bibfnamefont {J.~J.}\ \bibnamefont
  {Hopfield}},\ }\href {https://www.pnas.org/doi/abs/10.1073/pnas.79.8.2554}
  {\bibfield  {journal} {\bibinfo  {journal} {Proceedings of the national
  academy of sciences}\ }\textbf {\bibinfo {volume} {79}},\ \bibinfo {pages}
  {2554} (\bibinfo {year} {1982})}\BibitemShut {NoStop}%
\bibitem [{\citenamefont {Hinton}(2002)}]{hinton2002training}%
  \BibitemOpen
  \bibfield  {author} {\bibinfo {author} {\bibfnamefont {G.~E.}\ \bibnamefont
  {Hinton}},\ }\href {https://ieeexplore.ieee.org/abstract/document/6789337/}
  {\bibfield  {journal} {\bibinfo  {journal} {Neural computation}\ }\textbf
  {\bibinfo {volume} {14}},\ \bibinfo {pages} {1771} (\bibinfo {year}
  {2002})}\BibitemShut {NoStop}%
\bibitem [{\citenamefont {Carleo}\ and\ \citenamefont
  {Troyer}(2017)}]{carleo2017solving}%
  \BibitemOpen
  \bibfield  {author} {\bibinfo {author} {\bibfnamefont {G.}~\bibnamefont
  {Carleo}}\ and\ \bibinfo {author} {\bibfnamefont {M.}~\bibnamefont
  {Troyer}},\ }\href {https://www.science.org/doi/full/10.1126/science.aag2302}
  {\bibfield  {journal} {\bibinfo  {journal} {Science}\ }\textbf {\bibinfo
  {volume} {355}},\ \bibinfo {pages} {602} (\bibinfo {year}
  {2017})}\BibitemShut {NoStop}%
\bibitem [{\citenamefont {Cai}\ and\ \citenamefont
  {Liu}(2018)}]{cai2018approximating}%
  \BibitemOpen
  \bibfield  {author} {\bibinfo {author} {\bibfnamefont {Z.}~\bibnamefont
  {Cai}}\ and\ \bibinfo {author} {\bibfnamefont {J.}~\bibnamefont {Liu}},\
  }\href {https://journals.aps.org/prb/abstract/10.1103/PhysRevB.97.035116}
  {\bibfield  {journal} {\bibinfo  {journal} {Physical Review B}\ }\textbf
  {\bibinfo {volume} {97}},\ \bibinfo {pages} {035116} (\bibinfo {year}
  {2018})}\BibitemShut {NoStop}%
\bibitem [{\citenamefont {Choo}\ \emph {et~al.}(2018)\citenamefont {Choo},
  \citenamefont {Carleo}, \citenamefont {Regnault},\ and\ \citenamefont
  {Neupert}}]{choo2018symmetries}%
  \BibitemOpen
  \bibfield  {author} {\bibinfo {author} {\bibfnamefont {K.}~\bibnamefont
  {Choo}}, \bibinfo {author} {\bibfnamefont {G.}~\bibnamefont {Carleo}},
  \bibinfo {author} {\bibfnamefont {N.}~\bibnamefont {Regnault}},\ and\
  \bibinfo {author} {\bibfnamefont {T.}~\bibnamefont {Neupert}},\ }\href
  {https://journals.aps.org/prl/abstract/10.1103/PhysRevLett.121.167204}
  {\bibfield  {journal} {\bibinfo  {journal} {Physical review letters}\
  }\textbf {\bibinfo {volume} {121}},\ \bibinfo {pages} {167204} (\bibinfo
  {year} {2018})}\BibitemShut {NoStop}%
\bibitem [{\citenamefont {Liang}\ \emph {et~al.}(2018)\citenamefont {Liang},
  \citenamefont {Liu}, \citenamefont {Lin}, \citenamefont {Guo}, \citenamefont
  {Zhang},\ and\ \citenamefont {He}}]{liang2018solving}%
  \BibitemOpen
  \bibfield  {author} {\bibinfo {author} {\bibfnamefont {X.}~\bibnamefont
  {Liang}}, \bibinfo {author} {\bibfnamefont {W.-Y.}\ \bibnamefont {Liu}},
  \bibinfo {author} {\bibfnamefont {P.-Z.}\ \bibnamefont {Lin}}, \bibinfo
  {author} {\bibfnamefont {G.-C.}\ \bibnamefont {Guo}}, \bibinfo {author}
  {\bibfnamefont {Y.-S.}\ \bibnamefont {Zhang}},\ and\ \bibinfo {author}
  {\bibfnamefont {L.}~\bibnamefont {He}},\ }\href
  {https://journals.aps.org/prb/abstract/10.1103/PhysRevB.98.104426} {\bibfield
   {journal} {\bibinfo  {journal} {Physical Review B}\ }\textbf {\bibinfo
  {volume} {98}},\ \bibinfo {pages} {104426} (\bibinfo {year}
  {2018})}\BibitemShut {NoStop}%
\bibitem [{\citenamefont {Choo}\ \emph {et~al.}(2019)\citenamefont {Choo},
  \citenamefont {Neupert},\ and\ \citenamefont {Carleo}}]{choo2019two}%
  \BibitemOpen
  \bibfield  {author} {\bibinfo {author} {\bibfnamefont {K.}~\bibnamefont
  {Choo}}, \bibinfo {author} {\bibfnamefont {T.}~\bibnamefont {Neupert}},\ and\
  \bibinfo {author} {\bibfnamefont {G.}~\bibnamefont {Carleo}},\ }\href
  {https://journals.aps.org/prb/abstract/10.1103/PhysRevB.100.125124}
  {\bibfield  {journal} {\bibinfo  {journal} {Physical Review B}\ }\textbf
  {\bibinfo {volume} {100}},\ \bibinfo {pages} {125124} (\bibinfo {year}
  {2019})}\BibitemShut {NoStop}%
\bibitem [{\citenamefont {Hibat-Allah}\ \emph {et~al.}(2020)\citenamefont
  {Hibat-Allah}, \citenamefont {Ganahl}, \citenamefont {Hayward}, \citenamefont
  {Melko},\ and\ \citenamefont {Carrasquilla}}]{hibat2020recurrent}%
  \BibitemOpen
  \bibfield  {author} {\bibinfo {author} {\bibfnamefont {M.}~\bibnamefont
  {Hibat-Allah}}, \bibinfo {author} {\bibfnamefont {M.}~\bibnamefont {Ganahl}},
  \bibinfo {author} {\bibfnamefont {L.~E.}\ \bibnamefont {Hayward}}, \bibinfo
  {author} {\bibfnamefont {R.~G.}\ \bibnamefont {Melko}},\ and\ \bibinfo
  {author} {\bibfnamefont {J.}~\bibnamefont {Carrasquilla}},\ }\href
  {https://journals.aps.org/prresearch/abstract/10.1103/PhysRevResearch.2.023358}
  {\bibfield  {journal} {\bibinfo  {journal} {Physical Review Research}\
  }\textbf {\bibinfo {volume} {2}},\ \bibinfo {pages} {023358} (\bibinfo {year}
  {2020})}\BibitemShut {NoStop}%
\bibitem [{\citenamefont {Sharir}\ \emph {et~al.}(2020)\citenamefont {Sharir},
  \citenamefont {Levine}, \citenamefont {Wies}, \citenamefont {Carleo},\ and\
  \citenamefont {Shashua}}]{sharir2020deep}%
  \BibitemOpen
  \bibfield  {author} {\bibinfo {author} {\bibfnamefont {O.}~\bibnamefont
  {Sharir}}, \bibinfo {author} {\bibfnamefont {Y.}~\bibnamefont {Levine}},
  \bibinfo {author} {\bibfnamefont {N.}~\bibnamefont {Wies}}, \bibinfo {author}
  {\bibfnamefont {G.}~\bibnamefont {Carleo}},\ and\ \bibinfo {author}
  {\bibfnamefont {A.}~\bibnamefont {Shashua}},\ }\href
  {https://journals.aps.org/prl/abstract/10.1103/PhysRevLett.124.020503}
  {\bibfield  {journal} {\bibinfo  {journal} {Physical review letters}\
  }\textbf {\bibinfo {volume} {124}},\ \bibinfo {pages} {020503} (\bibinfo
  {year} {2020})}\BibitemShut {NoStop}%
\bibitem [{\citenamefont {Barrett}\ \emph {et~al.}(2022)\citenamefont
  {Barrett}, \citenamefont {Malyshev},\ and\ \citenamefont
  {Lvovsky}}]{barrett2022autoregressive}%
  \BibitemOpen
  \bibfield  {author} {\bibinfo {author} {\bibfnamefont {T.~D.}\ \bibnamefont
  {Barrett}}, \bibinfo {author} {\bibfnamefont {A.}~\bibnamefont {Malyshev}},\
  and\ \bibinfo {author} {\bibfnamefont {A.}~\bibnamefont {Lvovsky}},\ }\href
  {https://www.nature.com/articles/s42256-022-00461-z} {\bibfield  {journal}
  {\bibinfo  {journal} {Nature Machine Intelligence}\ }\textbf {\bibinfo
  {volume} {4}},\ \bibinfo {pages} {351} (\bibinfo {year} {2022})}\BibitemShut
  {NoStop}%
\bibitem [{\citenamefont {Luo}\ \emph {et~al.}(2021)\citenamefont {Luo},
  \citenamefont {Chen}, \citenamefont {Hu}, \citenamefont {Zhao}, \citenamefont
  {Hur},\ and\ \citenamefont {Clark}}]{luo2021gauge}%
  \BibitemOpen
  \bibfield  {author} {\bibinfo {author} {\bibfnamefont {D.}~\bibnamefont
  {Luo}}, \bibinfo {author} {\bibfnamefont {Z.}~\bibnamefont {Chen}}, \bibinfo
  {author} {\bibfnamefont {K.}~\bibnamefont {Hu}}, \bibinfo {author}
  {\bibfnamefont {Z.}~\bibnamefont {Zhao}}, \bibinfo {author} {\bibfnamefont
  {V.~M.}\ \bibnamefont {Hur}},\ and\ \bibinfo {author} {\bibfnamefont {B.~K.}\
  \bibnamefont {Clark}},\ }\href {https://arxiv.org/abs/2101.07243} {\bibfield
  {journal} {\bibinfo  {journal} {arXiv preprint arXiv:2101.07243}\ } (\bibinfo
  {year} {2021})}\BibitemShut {NoStop}%
\bibitem [{\citenamefont {Deng}\ \emph {et~al.}(2017)\citenamefont {Deng},
  \citenamefont {Li},\ and\ \citenamefont {Das~Sarma}}]{deng2017quantum}%
  \BibitemOpen
  \bibfield  {author} {\bibinfo {author} {\bibfnamefont {D.-L.}\ \bibnamefont
  {Deng}}, \bibinfo {author} {\bibfnamefont {X.}~\bibnamefont {Li}},\ and\
  \bibinfo {author} {\bibfnamefont {S.}~\bibnamefont {Das~Sarma}},\ }\href
  {https://journals.aps.org/prx/abstract/10.1103/PhysRevX.7.021021} {\bibfield
  {journal} {\bibinfo  {journal} {Physical Review X}\ }\textbf {\bibinfo
  {volume} {7}},\ \bibinfo {pages} {021021} (\bibinfo {year}
  {2017})}\BibitemShut {NoStop}%
\bibitem [{\citenamefont {Vaswani}\ \emph {et~al.}(2017)\citenamefont
  {Vaswani}, \citenamefont {Shazeer}, \citenamefont {Parmar}, \citenamefont
  {Uszkoreit}, \citenamefont {Jones}, \citenamefont {Gomez}, \citenamefont
  {Kaiser},\ and\ \citenamefont {Polosukhin}}]{vaswani2017attention}%
  \BibitemOpen
  \bibfield  {author} {\bibinfo {author} {\bibfnamefont {A.}~\bibnamefont
  {Vaswani}}, \bibinfo {author} {\bibfnamefont {N.}~\bibnamefont {Shazeer}},
  \bibinfo {author} {\bibfnamefont {N.}~\bibnamefont {Parmar}}, \bibinfo
  {author} {\bibfnamefont {J.}~\bibnamefont {Uszkoreit}}, \bibinfo {author}
  {\bibfnamefont {L.}~\bibnamefont {Jones}}, \bibinfo {author} {\bibfnamefont
  {A.~N.}\ \bibnamefont {Gomez}}, \bibinfo {author} {\bibfnamefont
  {{\L}.}~\bibnamefont {Kaiser}},\ and\ \bibinfo {author} {\bibfnamefont
  {I.}~\bibnamefont {Polosukhin}},\ }\href
  {https://proceedings.neurips.cc/paper/2017/hash/3f5ee243547dee91fbd053c1c4a845aa-Abstract.html}
  {\bibfield  {journal} {\bibinfo  {journal} {Advances in neural information
  processing systems}\ }\textbf {\bibinfo {volume} {30}} (\bibinfo {year}
  {2017})}\BibitemShut {NoStop}%
\bibitem [{\citenamefont {Devlin}\ \emph {et~al.}(2018)\citenamefont {Devlin},
  \citenamefont {Chang}, \citenamefont {Lee},\ and\ \citenamefont
  {Toutanova}}]{devlin2018bert}%
  \BibitemOpen
  \bibfield  {author} {\bibinfo {author} {\bibfnamefont {J.}~\bibnamefont
  {Devlin}}, \bibinfo {author} {\bibfnamefont {M.-W.}\ \bibnamefont {Chang}},
  \bibinfo {author} {\bibfnamefont {K.}~\bibnamefont {Lee}},\ and\ \bibinfo
  {author} {\bibfnamefont {K.}~\bibnamefont {Toutanova}},\ }\href
  {https://arxiv.org/abs/1810.04805} {\bibfield  {journal} {\bibinfo  {journal}
  {arXiv preprint arXiv:1810.04805}\ } (\bibinfo {year} {2018})}\BibitemShut
  {NoStop}%
\bibitem [{\citenamefont {Radford}\ \emph {et~al.}(2018)\citenamefont
  {Radford}, \citenamefont {Narasimhan}, \citenamefont {Salimans},
  \citenamefont {Sutskever} \emph {et~al.}}]{radford2018improving}%
  \BibitemOpen
  \bibfield  {author} {\bibinfo {author} {\bibfnamefont {A.}~\bibnamefont
  {Radford}}, \bibinfo {author} {\bibfnamefont {K.}~\bibnamefont {Narasimhan}},
  \bibinfo {author} {\bibfnamefont {T.}~\bibnamefont {Salimans}}, \bibinfo
  {author} {\bibfnamefont {I.}~\bibnamefont {Sutskever}}, \emph {et~al.},\
  }\href
  {https://www.cs.ubc.ca/~amuham01/LING530/papers/radford2018improving.pdf}
  {\bibfield  {journal} {\bibinfo  {journal} {OpenAI blog}\ } (\bibinfo {year}
  {2018})}\BibitemShut {NoStop}%
\bibitem [{\citenamefont {Radford}\ \emph {et~al.}(2019)\citenamefont
  {Radford}, \citenamefont {Wu}, \citenamefont {Child}, \citenamefont {Luan},
  \citenamefont {Amodei}, \citenamefont {Sutskever} \emph
  {et~al.}}]{radford2019language}%
  \BibitemOpen
  \bibfield  {author} {\bibinfo {author} {\bibfnamefont {A.}~\bibnamefont
  {Radford}}, \bibinfo {author} {\bibfnamefont {J.}~\bibnamefont {Wu}},
  \bibinfo {author} {\bibfnamefont {R.}~\bibnamefont {Child}}, \bibinfo
  {author} {\bibfnamefont {D.}~\bibnamefont {Luan}}, \bibinfo {author}
  {\bibfnamefont {D.}~\bibnamefont {Amodei}}, \bibinfo {author} {\bibfnamefont
  {I.}~\bibnamefont {Sutskever}}, \emph {et~al.},\ }\href
  {http://www.persagen.com/files/misc/radford2019language.pdf} {\bibfield
  {journal} {\bibinfo  {journal} {OpenAI blog}\ }\textbf {\bibinfo {volume}
  {1}},\ \bibinfo {pages} {9} (\bibinfo {year} {2019})}\BibitemShut {NoStop}%
\bibitem [{\citenamefont {Brown}\ \emph {et~al.}(2020)\citenamefont {Brown},
  \citenamefont {Mann}, \citenamefont {Ryder}, \citenamefont {Subbiah},
  \citenamefont {Kaplan}, \citenamefont {Dhariwal}, \citenamefont
  {Neelakantan}, \citenamefont {Shyam}, \citenamefont {Sastry}, \citenamefont
  {Askell} \emph {et~al.}}]{brown2020language}%
  \BibitemOpen
  \bibfield  {author} {\bibinfo {author} {\bibfnamefont {T.}~\bibnamefont
  {Brown}}, \bibinfo {author} {\bibfnamefont {B.}~\bibnamefont {Mann}},
  \bibinfo {author} {\bibfnamefont {N.}~\bibnamefont {Ryder}}, \bibinfo
  {author} {\bibfnamefont {M.}~\bibnamefont {Subbiah}}, \bibinfo {author}
  {\bibfnamefont {J.~D.}\ \bibnamefont {Kaplan}}, \bibinfo {author}
  {\bibfnamefont {P.}~\bibnamefont {Dhariwal}}, \bibinfo {author}
  {\bibfnamefont {A.}~\bibnamefont {Neelakantan}}, \bibinfo {author}
  {\bibfnamefont {P.}~\bibnamefont {Shyam}}, \bibinfo {author} {\bibfnamefont
  {G.}~\bibnamefont {Sastry}}, \bibinfo {author} {\bibfnamefont
  {A.}~\bibnamefont {Askell}}, \emph {et~al.},\ }\href
  {https://proceedings.neurips.cc/paper/2020/hash/1457c0d6bfcb4967418bfb8ac142f64a-Abstract.html}
  {\bibfield  {journal} {\bibinfo  {journal} {Advances in neural information
  processing systems}\ }\textbf {\bibinfo {volume} {33}},\ \bibinfo {pages}
  {1877} (\bibinfo {year} {2020})}\BibitemShut {NoStop}%
\bibitem [{\citenamefont {Dosovitskiy}\ \emph {et~al.}(2020)\citenamefont
  {Dosovitskiy}, \citenamefont {Beyer}, \citenamefont {Kolesnikov},
  \citenamefont {Weissenborn}, \citenamefont {Zhai}, \citenamefont
  {Unterthiner}, \citenamefont {Dehghani}, \citenamefont {Minderer},
  \citenamefont {Heigold}, \citenamefont {Gelly} \emph
  {et~al.}}]{dosovitskiy2020image}%
  \BibitemOpen
  \bibfield  {author} {\bibinfo {author} {\bibfnamefont {A.}~\bibnamefont
  {Dosovitskiy}}, \bibinfo {author} {\bibfnamefont {L.}~\bibnamefont {Beyer}},
  \bibinfo {author} {\bibfnamefont {A.}~\bibnamefont {Kolesnikov}}, \bibinfo
  {author} {\bibfnamefont {D.}~\bibnamefont {Weissenborn}}, \bibinfo {author}
  {\bibfnamefont {X.}~\bibnamefont {Zhai}}, \bibinfo {author} {\bibfnamefont
  {T.}~\bibnamefont {Unterthiner}}, \bibinfo {author} {\bibfnamefont
  {M.}~\bibnamefont {Dehghani}}, \bibinfo {author} {\bibfnamefont
  {M.}~\bibnamefont {Minderer}}, \bibinfo {author} {\bibfnamefont
  {G.}~\bibnamefont {Heigold}}, \bibinfo {author} {\bibfnamefont
  {S.}~\bibnamefont {Gelly}}, \emph {et~al.},\ }\href
  {https://arxiv.org/abs/2010.11929} {\bibfield  {journal} {\bibinfo  {journal}
  {arXiv preprint arXiv:2010.11929}\ } (\bibinfo {year} {2020})}\BibitemShut
  {NoStop}%
\bibitem [{\citenamefont {Dong}\ \emph {et~al.}(2018)\citenamefont {Dong},
  \citenamefont {Xu},\ and\ \citenamefont {Xu}}]{dong2018speech}%
  \BibitemOpen
  \bibfield  {author} {\bibinfo {author} {\bibfnamefont {L.}~\bibnamefont
  {Dong}}, \bibinfo {author} {\bibfnamefont {S.}~\bibnamefont {Xu}},\ and\
  \bibinfo {author} {\bibfnamefont {B.}~\bibnamefont {Xu}},\ }in\ \href
  {https://ieeexplore.ieee.org/abstract/document/8462506} {\emph {\bibinfo
  {booktitle} {2018 IEEE International Conference on Acoustics, Speech and
  Signal Processing (ICASSP)}}}\ (\bibinfo {organization} {IEEE},\ \bibinfo
  {year} {2018})\ pp.\ \bibinfo {pages} {5884--5888}\BibitemShut {NoStop}%
\bibitem [{\citenamefont {Veličković}\ \emph {et~al.}(2018)\citenamefont
  {Veličković}, \citenamefont {Cucurull}, \citenamefont {Casanova},
  \citenamefont {Romero}, \citenamefont {Liò},\ and\ \citenamefont
  {Bengio}}]{velickovic2018graph}%
  \BibitemOpen
  \bibfield  {author} {\bibinfo {author} {\bibfnamefont {P.}~\bibnamefont
  {Veličković}}, \bibinfo {author} {\bibfnamefont {G.}~\bibnamefont
  {Cucurull}}, \bibinfo {author} {\bibfnamefont {A.}~\bibnamefont {Casanova}},
  \bibinfo {author} {\bibfnamefont {A.}~\bibnamefont {Romero}}, \bibinfo
  {author} {\bibfnamefont {P.}~\bibnamefont {Liò}},\ and\ \bibinfo {author}
  {\bibfnamefont {Y.}~\bibnamefont {Bengio}},\ }in\ \href
  {https://openreview.net/forum?id=rJXMpikCZ} {\emph {\bibinfo {booktitle}
  {International Conference on Learning Representations}}}\ (\bibinfo {year}
  {2018})\BibitemShut {NoStop}%
\bibitem [{\citenamefont {Reed}\ \emph {et~al.}(2022)\citenamefont {Reed},
  \citenamefont {Zolna}, \citenamefont {Parisotto}, \citenamefont
  {Colmenarejo}, \citenamefont {Novikov}, \citenamefont {Barth-Maron},
  \citenamefont {Gimenez}, \citenamefont {Sulsky}, \citenamefont {Kay},
  \citenamefont {Springenberg} \emph {et~al.}}]{reed2022generalist}%
  \BibitemOpen
  \bibfield  {author} {\bibinfo {author} {\bibfnamefont {S.}~\bibnamefont
  {Reed}}, \bibinfo {author} {\bibfnamefont {K.}~\bibnamefont {Zolna}},
  \bibinfo {author} {\bibfnamefont {E.}~\bibnamefont {Parisotto}}, \bibinfo
  {author} {\bibfnamefont {S.~G.}\ \bibnamefont {Colmenarejo}}, \bibinfo
  {author} {\bibfnamefont {A.}~\bibnamefont {Novikov}}, \bibinfo {author}
  {\bibfnamefont {G.}~\bibnamefont {Barth-Maron}}, \bibinfo {author}
  {\bibfnamefont {M.}~\bibnamefont {Gimenez}}, \bibinfo {author} {\bibfnamefont
  {Y.}~\bibnamefont {Sulsky}}, \bibinfo {author} {\bibfnamefont
  {J.}~\bibnamefont {Kay}}, \bibinfo {author} {\bibfnamefont {J.~T.}\
  \bibnamefont {Springenberg}}, \emph {et~al.},\ }\href
  {https://arxiv.org/abs/2205.06175} {\bibfield  {journal} {\bibinfo  {journal}
  {arXiv preprint arXiv:2205.06175}\ } (\bibinfo {year} {2022})}\BibitemShut
  {NoStop}%
\bibitem [{\citenamefont {Luo}\ \emph {et~al.}(2022)\citenamefont {Luo},
  \citenamefont {Chen}, \citenamefont {Carrasquilla},\ and\ \citenamefont
  {Clark}}]{luo2022autoregressive}%
  \BibitemOpen
  \bibfield  {author} {\bibinfo {author} {\bibfnamefont {D.}~\bibnamefont
  {Luo}}, \bibinfo {author} {\bibfnamefont {Z.}~\bibnamefont {Chen}}, \bibinfo
  {author} {\bibfnamefont {J.}~\bibnamefont {Carrasquilla}},\ and\ \bibinfo
  {author} {\bibfnamefont {B.~K.}\ \bibnamefont {Clark}},\ }\href
  {https://journals.aps.org/prl/abstract/10.1103/PhysRevLett.128.090501}
  {\bibfield  {journal} {\bibinfo  {journal} {Physical review letters}\
  }\textbf {\bibinfo {volume} {128}},\ \bibinfo {pages} {090501} (\bibinfo
  {year} {2022})}\BibitemShut {NoStop}%
\bibitem [{\citenamefont {Cha}\ \emph {et~al.}(2021)\citenamefont {Cha},
  \citenamefont {Ginsparg}, \citenamefont {Wu}, \citenamefont {Carrasquilla},
  \citenamefont {McMahon},\ and\ \citenamefont {Kim}}]{cha2021attention}%
  \BibitemOpen
  \bibfield  {author} {\bibinfo {author} {\bibfnamefont {P.}~\bibnamefont
  {Cha}}, \bibinfo {author} {\bibfnamefont {P.}~\bibnamefont {Ginsparg}},
  \bibinfo {author} {\bibfnamefont {F.}~\bibnamefont {Wu}}, \bibinfo {author}
  {\bibfnamefont {J.}~\bibnamefont {Carrasquilla}}, \bibinfo {author}
  {\bibfnamefont {P.~L.}\ \bibnamefont {McMahon}},\ and\ \bibinfo {author}
  {\bibfnamefont {E.-A.}\ \bibnamefont {Kim}},\ }\href
  {https://iopscience.iop.org/article/10.1088/2632-2153/ac362b/meta} {\bibfield
   {journal} {\bibinfo  {journal} {Machine Learning: Science and Technology}\
  }\textbf {\bibinfo {volume} {3}},\ \bibinfo {pages} {01LT01} (\bibinfo {year}
  {2021})}\BibitemShut {NoStop}%
\bibitem [{\citenamefont {Carrasquilla}\ \emph {et~al.}(2021)\citenamefont
  {Carrasquilla}, \citenamefont {Luo}, \citenamefont {P{\'e}rez}, \citenamefont
  {Milsted}, \citenamefont {Clark}, \citenamefont {Volkovs},\ and\
  \citenamefont {Aolita}}]{carrasquilla2021probabilistic}%
  \BibitemOpen
  \bibfield  {author} {\bibinfo {author} {\bibfnamefont {J.}~\bibnamefont
  {Carrasquilla}}, \bibinfo {author} {\bibfnamefont {D.}~\bibnamefont {Luo}},
  \bibinfo {author} {\bibfnamefont {F.}~\bibnamefont {P{\'e}rez}}, \bibinfo
  {author} {\bibfnamefont {A.}~\bibnamefont {Milsted}}, \bibinfo {author}
  {\bibfnamefont {B.~K.}\ \bibnamefont {Clark}}, \bibinfo {author}
  {\bibfnamefont {M.}~\bibnamefont {Volkovs}},\ and\ \bibinfo {author}
  {\bibfnamefont {L.}~\bibnamefont {Aolita}},\ }\href
  {https://journals.aps.org/pra/abstract/10.1103/PhysRevA.104.032610}
  {\bibfield  {journal} {\bibinfo  {journal} {Physical Review A}\ }\textbf
  {\bibinfo {volume} {104}},\ \bibinfo {pages} {032610} (\bibinfo {year}
  {2021})}\BibitemShut {NoStop}%
\bibitem [{\citenamefont {Zhang}\ \emph {et~al.}(2020)\citenamefont {Zhang},
  \citenamefont {Zheng}, \citenamefont {Zhang},\ and\ \citenamefont
  {Deng}}]{zhang2020topological}%
  \BibitemOpen
  \bibfield  {author} {\bibinfo {author} {\bibfnamefont {Y.-H.}\ \bibnamefont
  {Zhang}}, \bibinfo {author} {\bibfnamefont {P.-L.}\ \bibnamefont {Zheng}},
  \bibinfo {author} {\bibfnamefont {Y.}~\bibnamefont {Zhang}},\ and\ \bibinfo
  {author} {\bibfnamefont {D.-L.}\ \bibnamefont {Deng}},\ }\href
  {https://journals.aps.org/prl/abstract/10.1103/PhysRevLett.125.170501}
  {\bibfield  {journal} {\bibinfo  {journal} {Physical Review Letters}\
  }\textbf {\bibinfo {volume} {125}},\ \bibinfo {pages} {170501} (\bibinfo
  {year} {2020})}\BibitemShut {NoStop}%
\bibitem [{\citenamefont {Carrasquilla}\ and\ \citenamefont
  {Melko}(2017)}]{carrasquilla2017machine}%
  \BibitemOpen
  \bibfield  {author} {\bibinfo {author} {\bibfnamefont {J.}~\bibnamefont
  {Carrasquilla}}\ and\ \bibinfo {author} {\bibfnamefont {R.~G.}\ \bibnamefont
  {Melko}},\ }\href {https://www.nature.com/articles/nphys4035} {\bibfield
  {journal} {\bibinfo  {journal} {Nature Physics}\ }\textbf {\bibinfo {volume}
  {13}},\ \bibinfo {pages} {431} (\bibinfo {year} {2017})}\BibitemShut
  {NoStop}%
\bibitem [{\citenamefont {Van~Nieuwenburg}\ \emph {et~al.}(2017)\citenamefont
  {Van~Nieuwenburg}, \citenamefont {Liu},\ and\ \citenamefont
  {Huber}}]{van2017learning}%
  \BibitemOpen
  \bibfield  {author} {\bibinfo {author} {\bibfnamefont {E.~P.}\ \bibnamefont
  {Van~Nieuwenburg}}, \bibinfo {author} {\bibfnamefont {Y.-H.}\ \bibnamefont
  {Liu}},\ and\ \bibinfo {author} {\bibfnamefont {S.~D.}\ \bibnamefont
  {Huber}},\ }\href {https://www.nature.com/articles/nphys4037} {\bibfield
  {journal} {\bibinfo  {journal} {Nature Physics}\ }\textbf {\bibinfo {volume}
  {13}},\ \bibinfo {pages} {435} (\bibinfo {year} {2017})}\BibitemShut
  {NoStop}%
\bibitem [{\citenamefont {Torlai}\ \emph {et~al.}(2018)\citenamefont {Torlai},
  \citenamefont {Mazzola}, \citenamefont {Carrasquilla}, \citenamefont
  {Troyer}, \citenamefont {Melko},\ and\ \citenamefont
  {Carleo}}]{torlai2018neural}%
  \BibitemOpen
  \bibfield  {author} {\bibinfo {author} {\bibfnamefont {G.}~\bibnamefont
  {Torlai}}, \bibinfo {author} {\bibfnamefont {G.}~\bibnamefont {Mazzola}},
  \bibinfo {author} {\bibfnamefont {J.}~\bibnamefont {Carrasquilla}}, \bibinfo
  {author} {\bibfnamefont {M.}~\bibnamefont {Troyer}}, \bibinfo {author}
  {\bibfnamefont {R.}~\bibnamefont {Melko}},\ and\ \bibinfo {author}
  {\bibfnamefont {G.}~\bibnamefont {Carleo}},\ }\href
  {https://www.nature.com/articles/s41567-018-0048-5} {\bibfield  {journal}
  {\bibinfo  {journal} {Nature Physics}\ }\textbf {\bibinfo {volume} {14}},\
  \bibinfo {pages} {447} (\bibinfo {year} {2018})}\BibitemShut {NoStop}%
\bibitem [{\citenamefont {Zhang}\ and\ \citenamefont {{Di
  Ventra}}(2022)}]{zhang2021efficient}%
  \BibitemOpen
  \bibfield  {author} {\bibinfo {author} {\bibfnamefont {Y.-H.}\ \bibnamefont
  {Zhang}}\ and\ \bibinfo {author} {\bibfnamefont {M.}~\bibnamefont {{Di
  Ventra}}},\ }\href
  {https://journals.aps.org/pra/abstract/10.1103/PhysRevA.106.042420}
  {\bibfield  {journal} {\bibinfo  {journal} {Physical Review A}\ }\textbf
  {\bibinfo {volume} {106}},\ \bibinfo {pages} {042420} (\bibinfo {year}
  {2022})}\BibitemShut {NoStop}%
\bibitem [{\citenamefont {Schulz}\ \emph {et~al.}(1996)\citenamefont {Schulz},
  \citenamefont {Ziman},\ and\ \citenamefont {Poilblanc}}]{schulz1996magnetic}%
  \BibitemOpen
  \bibfield  {author} {\bibinfo {author} {\bibfnamefont {H.}~\bibnamefont
  {Schulz}}, \bibinfo {author} {\bibfnamefont {T.}~\bibnamefont {Ziman}},\ and\
  \bibinfo {author} {\bibfnamefont {D.}~\bibnamefont {Poilblanc}},\ }\href
  {https://jp1.journaldephysique.org/articles/jp1/abs/1996/05/jp1v6p675/jp1v6p675.html}
  {\bibfield  {journal} {\bibinfo  {journal} {Journal de Physique I}\ }\textbf
  {\bibinfo {volume} {6}},\ \bibinfo {pages} {675} (\bibinfo {year}
  {1996})}\BibitemShut {NoStop}%
\bibitem [{\citenamefont {Myung}(2003)}]{myung2003tutorial}%
  \BibitemOpen
  \bibfield  {author} {\bibinfo {author} {\bibfnamefont {I.~J.}\ \bibnamefont
  {Myung}},\ }\href
  {https://www.sciencedirect.com/science/article/pii/S0022249602000287}
  {\bibfield  {journal} {\bibinfo  {journal} {Journal of mathematical
  Psychology}\ }\textbf {\bibinfo {volume} {47}},\ \bibinfo {pages} {90}
  (\bibinfo {year} {2003})}\BibitemShut {NoStop}%
\bibitem [{\citenamefont {Aaronson}(2019)}]{aaronson2019shadow}%
  \BibitemOpen
  \bibfield  {author} {\bibinfo {author} {\bibfnamefont {S.}~\bibnamefont
  {Aaronson}},\ }\href {https://epubs.siam.org/doi/abs/10.1137/18M120275X}
  {\bibfield  {journal} {\bibinfo  {journal} {SIAM Journal on Computing}\
  }\textbf {\bibinfo {volume} {49}},\ \bibinfo {pages} {STOC18} (\bibinfo
  {year} {2019})}\BibitemShut {NoStop}%
\bibitem [{\citenamefont {Huang}\ \emph {et~al.}(2020)\citenamefont {Huang},
  \citenamefont {Kueng},\ and\ \citenamefont {Preskill}}]{huang2020predicting}%
  \BibitemOpen
  \bibfield  {author} {\bibinfo {author} {\bibfnamefont {H.-Y.}\ \bibnamefont
  {Huang}}, \bibinfo {author} {\bibfnamefont {R.}~\bibnamefont {Kueng}},\ and\
  \bibinfo {author} {\bibfnamefont {J.}~\bibnamefont {Preskill}},\ }\href
  {https://www.nature.com/articles/s41567-020-0932-7} {\bibfield  {journal}
  {\bibinfo  {journal} {Nature Physics}\ }\textbf {\bibinfo {volume} {16}},\
  \bibinfo {pages} {1050} (\bibinfo {year} {2020})}\BibitemShut {NoStop}%
\bibitem [{\citenamefont {Binder}(1981)}]{binder1981finite}%
  \BibitemOpen
  \bibfield  {author} {\bibinfo {author} {\bibfnamefont {K.}~\bibnamefont
  {Binder}},\ }\href {https://link.springer.com/article/10.1007/BF01293604}
  {\bibfield  {journal} {\bibinfo  {journal} {Zeitschrift f{\"u}r Physik B
  Condensed Matter}\ }\textbf {\bibinfo {volume} {43}},\ \bibinfo {pages} {119}
  (\bibinfo {year} {1981})}\BibitemShut {NoStop}%
\bibitem [{\citenamefont {Pang}\ \emph {et~al.}(2019)\citenamefont {Pang},
  \citenamefont {Muniandy},\ and\ \citenamefont {Kamali}}]{pang2019critical}%
  \BibitemOpen
  \bibfield  {author} {\bibinfo {author} {\bibfnamefont {S.~Y.}\ \bibnamefont
  {Pang}}, \bibinfo {author} {\bibfnamefont {S.~V.}\ \bibnamefont {Muniandy}},\
  and\ \bibinfo {author} {\bibfnamefont {M.~Z.~M.}\ \bibnamefont {Kamali}},\
  }\href {https://link.springer.com/article/10.1007/s10773-019-04279-1}
  {\bibfield  {journal} {\bibinfo  {journal} {International Journal of
  Theoretical Physics}\ }\textbf {\bibinfo {volume} {58}},\ \bibinfo {pages}
  {4139} (\bibinfo {year} {2019})}\BibitemShut {NoStop}%
\bibitem [{\citenamefont {Suzuki}\ \emph {et~al.}(2012)\citenamefont {Suzuki},
  \citenamefont {Inoue},\ and\ \citenamefont
  {Chakrabarti}}]{suzuki2012quantum}%
  \BibitemOpen
  \bibfield  {author} {\bibinfo {author} {\bibfnamefont {S.}~\bibnamefont
  {Suzuki}}, \bibinfo {author} {\bibfnamefont {J.-i.}\ \bibnamefont {Inoue}},\
  and\ \bibinfo {author} {\bibfnamefont {B.~K.}\ \bibnamefont {Chakrabarti}},\
  }\href {https://link.springer.com/book/10.1007/978-3-642-33039-1} {\emph
  {\bibinfo {title} {Quantum Ising phases and transitions in transverse Ising
  models}}},\ Vol.\ \bibinfo {volume} {862}\ (\bibinfo  {publisher}
  {Springer},\ \bibinfo {year} {2012})\BibitemShut {NoStop}%
\bibitem [{\citenamefont {Hauschild}\ and\ \citenamefont
  {Pollmann}(2018)}]{tenpy}%
  \BibitemOpen
  \bibfield  {author} {\bibinfo {author} {\bibfnamefont {J.}~\bibnamefont
  {Hauschild}}\ and\ \bibinfo {author} {\bibfnamefont {F.}~\bibnamefont
  {Pollmann}},\ }\href {https://doi.org/10.21468/SciPostPhysLectNotes.5}
  {\bibfield  {journal} {\bibinfo  {journal} {SciPost Phys. Lect. Notes}\ ,\
  \bibinfo {pages} {5}} (\bibinfo {year} {2018})},\ \bibinfo {note} {code
  available from \url{https://github.com/tenpy/tenpy}},\ \Eprint
  {https://arxiv.org/abs/1805.00055} {arXiv:1805.00055} \BibitemShut {NoStop}%
\bibitem [{\citenamefont {Di~Ventra}(2022)}]{di2022memcomputing}%
  \BibitemOpen
  \bibfield  {author} {\bibinfo {author} {\bibfnamefont {M.}~\bibnamefont
  {Di~Ventra}},\ }\href {https://academic.oup.com/book/42003} {\emph {\bibinfo
  {title} {MemComputing: Fundamentals and Applications}}}\ (\bibinfo
  {publisher} {Oxford University Press},\ \bibinfo {year} {2022})\BibitemShut
  {NoStop}%
\bibitem [{\citenamefont {Zheng}\ and\ \citenamefont
  {Casari}(2018)}]{zheng2018feature}%
  \BibitemOpen
  \bibfield  {author} {\bibinfo {author} {\bibfnamefont {A.}~\bibnamefont
  {Zheng}}\ and\ \bibinfo {author} {\bibfnamefont {A.}~\bibnamefont {Casari}},\
  }\href
  {https://www.oreilly.com/library/view/feature-engineering-for/9781491953235/}
  {\emph {\bibinfo {title} {Feature engineering for machine learning:
  principles and techniques for data scientists}}}\ (\bibinfo  {publisher} {"
  O'Reilly Media, Inc."},\ \bibinfo {year} {2018})\BibitemShut {NoStop}%
\bibitem [{\citenamefont {Wang}\ and\ \citenamefont
  {Liu}(2019)}]{wang2019translating}%
  \BibitemOpen
  \bibfield  {author} {\bibinfo {author} {\bibfnamefont {Z.}~\bibnamefont
  {Wang}}\ and\ \bibinfo {author} {\bibfnamefont {J.-C.}\ \bibnamefont {Liu}},\
  }\href@noop {} {\bibinfo {title} {Translating math formula images to latex
  sequences using deep neural networks with sequence-level training}} (\bibinfo
  {year} {2019}),\ \Eprint {https://arxiv.org/abs/1908.11415} {arXiv:1908.11415
  [cs.LG]} \BibitemShut {NoStop}%
\bibitem [{\citenamefont {Nair}\ and\ \citenamefont
  {Hinton}(2010)}]{nair2010rectified}%
  \BibitemOpen
  \bibfield  {author} {\bibinfo {author} {\bibfnamefont {V.}~\bibnamefont
  {Nair}}\ and\ \bibinfo {author} {\bibfnamefont {G.~E.}\ \bibnamefont
  {Hinton}},\ }in\ \href {https://openreview.net/forum?id=rkb15iZdZB} {\emph
  {\bibinfo {booktitle} {Icml}}}\ (\bibinfo {year} {2010})\BibitemShut
  {NoStop}%
\bibitem [{\citenamefont {Bridle}(1990)}]{bridle1990probabilistic}%
  \BibitemOpen
  \bibfield  {author} {\bibinfo {author} {\bibfnamefont {J.~S.}\ \bibnamefont
  {Bridle}},\ }in\ \href
  {https://link.springer.com/chapter/10.1007/978-3-642-76153-9_28} {\emph
  {\bibinfo {booktitle} {Neurocomputing}}}\ (\bibinfo  {publisher} {Springer},\
  \bibinfo {year} {1990})\ pp.\ \bibinfo {pages} {227--236}\BibitemShut
  {NoStop}%
\bibitem [{\citenamefont {Paszke}\ \emph {et~al.}(2019)\citenamefont {Paszke},
  \citenamefont {Gross}, \citenamefont {Massa}, \citenamefont {Lerer},
  \citenamefont {Bradbury}, \citenamefont {Chanan}, \citenamefont {Killeen},
  \citenamefont {Lin}, \citenamefont {Gimelshein}, \citenamefont {Antiga} \emph
  {et~al.}}]{paszke2019pytorch}%
  \BibitemOpen
  \bibfield  {author} {\bibinfo {author} {\bibfnamefont {A.}~\bibnamefont
  {Paszke}}, \bibinfo {author} {\bibfnamefont {S.}~\bibnamefont {Gross}},
  \bibinfo {author} {\bibfnamefont {F.}~\bibnamefont {Massa}}, \bibinfo
  {author} {\bibfnamefont {A.}~\bibnamefont {Lerer}}, \bibinfo {author}
  {\bibfnamefont {J.}~\bibnamefont {Bradbury}}, \bibinfo {author}
  {\bibfnamefont {G.}~\bibnamefont {Chanan}}, \bibinfo {author} {\bibfnamefont
  {T.}~\bibnamefont {Killeen}}, \bibinfo {author} {\bibfnamefont
  {Z.}~\bibnamefont {Lin}}, \bibinfo {author} {\bibfnamefont {N.}~\bibnamefont
  {Gimelshein}}, \bibinfo {author} {\bibfnamefont {L.}~\bibnamefont {Antiga}},
  \emph {et~al.},\ }\href
  {https://proceedings.neurips.cc/paper/2019/hash/bdbca288fee7f92f2bfa9f7012727740-Abstract.html}
  {\bibfield  {journal} {\bibinfo  {journal} {Advances in neural information
  processing systems}\ }\textbf {\bibinfo {volume} {32}} (\bibinfo {year}
  {2019})}\BibitemShut {NoStop}%
\bibitem [{\citenamefont {Kingma}\ and\ \citenamefont
  {Ba}(2014)}]{kingma2014adam}%
  \BibitemOpen
  \bibfield  {author} {\bibinfo {author} {\bibfnamefont {D.~P.}\ \bibnamefont
  {Kingma}}\ and\ \bibinfo {author} {\bibfnamefont {J.}~\bibnamefont {Ba}},\
  }\href {https://arxiv.org/abs/1412.6980} {\bibfield  {journal} {\bibinfo
  {journal} {arXiv preprint arXiv:1412.6980}\ } (\bibinfo {year}
  {2014})}\BibitemShut {NoStop}%
\bibitem [{\citenamefont {Nelder}\ and\ \citenamefont
  {Mead}(1965)}]{nelder1965simplex}%
  \BibitemOpen
  \bibfield  {author} {\bibinfo {author} {\bibfnamefont {J.~A.}\ \bibnamefont
  {Nelder}}\ and\ \bibinfo {author} {\bibfnamefont {R.}~\bibnamefont {Mead}},\
  }\href {https://academic.oup.com/comjnl/article-abstract/7/4/308/354237}
  {\bibfield  {journal} {\bibinfo  {journal} {The computer journal}\ }\textbf
  {\bibinfo {volume} {7}},\ \bibinfo {pages} {308} (\bibinfo {year}
  {1965})}\BibitemShut {NoStop}%
\bibitem [{\citenamefont {Virtanen}\ \emph {et~al.}(2020)\citenamefont
  {Virtanen}, \citenamefont {Gommers}, \citenamefont {Oliphant}, \citenamefont
  {Haberland}, \citenamefont {Reddy}, \citenamefont {Cournapeau}, \citenamefont
  {Burovski}, \citenamefont {Peterson}, \citenamefont {Weckesser},
  \citenamefont {Bright}, \citenamefont {{van der Walt}}, \citenamefont
  {Brett}, \citenamefont {Wilson}, \citenamefont {Millman}, \citenamefont
  {Mayorov}, \citenamefont {Nelson}, \citenamefont {Jones}, \citenamefont
  {Kern}, \citenamefont {Larson}, \citenamefont {Carey}, \citenamefont {Polat},
  \citenamefont {Feng}, \citenamefont {Moore}, \citenamefont {{VanderPlas}},
  \citenamefont {Laxalde}, \citenamefont {Perktold}, \citenamefont {Cimrman},
  \citenamefont {Henriksen}, \citenamefont {Quintero}, \citenamefont {Harris},
  \citenamefont {Archibald}, \citenamefont {Ribeiro}, \citenamefont
  {Pedregosa}, \citenamefont {{van Mulbregt}},\ and\ \citenamefont {{SciPy 1.0
  Contributors}}}]{2020SciPy-NMeth}%
  \BibitemOpen
  \bibfield  {author} {\bibinfo {author} {\bibfnamefont {P.}~\bibnamefont
  {Virtanen}}, \bibinfo {author} {\bibfnamefont {R.}~\bibnamefont {Gommers}},
  \bibinfo {author} {\bibfnamefont {T.~E.}\ \bibnamefont {Oliphant}}, \bibinfo
  {author} {\bibfnamefont {M.}~\bibnamefont {Haberland}}, \bibinfo {author}
  {\bibfnamefont {T.}~\bibnamefont {Reddy}}, \bibinfo {author} {\bibfnamefont
  {D.}~\bibnamefont {Cournapeau}}, \bibinfo {author} {\bibfnamefont
  {E.}~\bibnamefont {Burovski}}, \bibinfo {author} {\bibfnamefont
  {P.}~\bibnamefont {Peterson}}, \bibinfo {author} {\bibfnamefont
  {W.}~\bibnamefont {Weckesser}}, \bibinfo {author} {\bibfnamefont
  {J.}~\bibnamefont {Bright}}, \bibinfo {author} {\bibfnamefont {S.~J.}\
  \bibnamefont {{van der Walt}}}, \bibinfo {author} {\bibfnamefont
  {M.}~\bibnamefont {Brett}}, \bibinfo {author} {\bibfnamefont
  {J.}~\bibnamefont {Wilson}}, \bibinfo {author} {\bibfnamefont {K.~J.}\
  \bibnamefont {Millman}}, \bibinfo {author} {\bibfnamefont {N.}~\bibnamefont
  {Mayorov}}, \bibinfo {author} {\bibfnamefont {A.~R.~J.}\ \bibnamefont
  {Nelson}}, \bibinfo {author} {\bibfnamefont {E.}~\bibnamefont {Jones}},
  \bibinfo {author} {\bibfnamefont {R.}~\bibnamefont {Kern}}, \bibinfo {author}
  {\bibfnamefont {E.}~\bibnamefont {Larson}}, \bibinfo {author} {\bibfnamefont
  {C.~J.}\ \bibnamefont {Carey}}, \bibinfo {author} {\bibfnamefont
  {{\.I}.}~\bibnamefont {Polat}}, \bibinfo {author} {\bibfnamefont
  {Y.}~\bibnamefont {Feng}}, \bibinfo {author} {\bibfnamefont {E.~W.}\
  \bibnamefont {Moore}}, \bibinfo {author} {\bibfnamefont {J.}~\bibnamefont
  {{VanderPlas}}}, \bibinfo {author} {\bibfnamefont {D.}~\bibnamefont
  {Laxalde}}, \bibinfo {author} {\bibfnamefont {J.}~\bibnamefont {Perktold}},
  \bibinfo {author} {\bibfnamefont {R.}~\bibnamefont {Cimrman}}, \bibinfo
  {author} {\bibfnamefont {I.}~\bibnamefont {Henriksen}}, \bibinfo {author}
  {\bibfnamefont {E.~A.}\ \bibnamefont {Quintero}}, \bibinfo {author}
  {\bibfnamefont {C.~R.}\ \bibnamefont {Harris}}, \bibinfo {author}
  {\bibfnamefont {A.~M.}\ \bibnamefont {Archibald}}, \bibinfo {author}
  {\bibfnamefont {A.~H.}\ \bibnamefont {Ribeiro}}, \bibinfo {author}
  {\bibfnamefont {F.}~\bibnamefont {Pedregosa}}, \bibinfo {author}
  {\bibfnamefont {P.}~\bibnamefont {{van Mulbregt}}},\ and\ \bibinfo {author}
  {\bibnamefont {{SciPy 1.0 Contributors}}},\ }\href
  {https://doi.org/10.1038/s41592-019-0686-2} {\bibfield  {journal} {\bibinfo
  {journal} {Nature Methods}\ }\textbf {\bibinfo {volume} {17}},\ \bibinfo
  {pages} {261} (\bibinfo {year} {2020})}\BibitemShut {NoStop}%
\bibitem [{\citenamefont {Carrasquilla}\ \emph {et~al.}(2019)\citenamefont
  {Carrasquilla}, \citenamefont {Torlai}, \citenamefont {Melko},\ and\
  \citenamefont {Aolita}}]{carrasquilla2019reconstructing}%
  \BibitemOpen
  \bibfield  {author} {\bibinfo {author} {\bibfnamefont {J.}~\bibnamefont
  {Carrasquilla}}, \bibinfo {author} {\bibfnamefont {G.}~\bibnamefont
  {Torlai}}, \bibinfo {author} {\bibfnamefont {R.~G.}\ \bibnamefont {Melko}},\
  and\ \bibinfo {author} {\bibfnamefont {L.}~\bibnamefont {Aolita}},\ }\href
  {https://www.nature.com/articles/s42256-019-0028-1} {\bibfield  {journal}
  {\bibinfo  {journal} {Nature Machine Intelligence}\ }\textbf {\bibinfo
  {volume} {1}},\ \bibinfo {pages} {155} (\bibinfo {year} {2019})}\BibitemShut
  {NoStop}%
\bibitem [{\citenamefont {Sorella}\ \emph {et~al.}(2007)\citenamefont
  {Sorella}, \citenamefont {Casula},\ and\ \citenamefont
  {Rocca}}]{sorella2007weak}%
  \BibitemOpen
  \bibfield  {author} {\bibinfo {author} {\bibfnamefont {S.}~\bibnamefont
  {Sorella}}, \bibinfo {author} {\bibfnamefont {M.}~\bibnamefont {Casula}},\
  and\ \bibinfo {author} {\bibfnamefont {D.}~\bibnamefont {Rocca}},\ }\href
  {https://aip.scitation.org/doi/10.1063/1.2746035} {\bibfield  {journal}
  {\bibinfo  {journal} {The Journal of chemical physics}\ }\textbf {\bibinfo
  {volume} {127}},\ \bibinfo {pages} {014105} (\bibinfo {year}
  {2007})}\BibitemShut {NoStop}%
\bibitem [{\citenamefont {Carreira-Perpinan}\ and\ \citenamefont
  {Hinton}(2005)}]{carreira2005contrastive}%
  \BibitemOpen
  \bibfield  {author} {\bibinfo {author} {\bibfnamefont {M.~A.}\ \bibnamefont
  {Carreira-Perpinan}}\ and\ \bibinfo {author} {\bibfnamefont {G.}~\bibnamefont
  {Hinton}},\ }in\ \href
  {https://proceedings.mlr.press/r5/carreira-perpinan05a.html} {\emph {\bibinfo
  {booktitle} {International workshop on artificial intelligence and
  statistics}}}\ (\bibinfo {organization} {PMLR},\ \bibinfo {year} {2005})\
  pp.\ \bibinfo {pages} {33--40}\BibitemShut {NoStop}%
\bibitem [{\citenamefont {Fisher}\ and\ \citenamefont
  {Barber}(1972)}]{fisher1972scaling}%
  \BibitemOpen
  \bibfield  {author} {\bibinfo {author} {\bibfnamefont {M.~E.}\ \bibnamefont
  {Fisher}}\ and\ \bibinfo {author} {\bibfnamefont {M.~N.}\ \bibnamefont
  {Barber}},\ }\href
  {https://journals.aps.org/prl/abstract/10.1103/PhysRevLett.28.1516}
  {\bibfield  {journal} {\bibinfo  {journal} {Physical Review Letters}\
  }\textbf {\bibinfo {volume} {28}},\ \bibinfo {pages} {1516} (\bibinfo {year}
  {1972})}\BibitemShut {NoStop}%
\bibitem [{\citenamefont {Schollw{\"o}ck}\ \emph {et~al.}(2008)\citenamefont
  {Schollw{\"o}ck}, \citenamefont {Richter}, \citenamefont {Farnell},\ and\
  \citenamefont {Bishop}}]{schollwock2008quantum}%
  \BibitemOpen
  \bibfield  {author} {\bibinfo {author} {\bibfnamefont {U.}~\bibnamefont
  {Schollw{\"o}ck}}, \bibinfo {author} {\bibfnamefont {J.}~\bibnamefont
  {Richter}}, \bibinfo {author} {\bibfnamefont {D.~J.}\ \bibnamefont
  {Farnell}},\ and\ \bibinfo {author} {\bibfnamefont {R.~F.}\ \bibnamefont
  {Bishop}},\ }\href {https://link.springer.com/book/10.1007/b96825} {\emph
  {\bibinfo {title} {Quantum magnetism}}},\ Vol.\ \bibinfo {volume} {645}\
  (\bibinfo  {publisher} {Springer},\ \bibinfo {year} {2008})\BibitemShut
  {NoStop}%
\end{thebibliography}%

\clearpage
\newpage
\appendix

% \begin{center} 
% {\large \bf Supplemental Material: Transformer Quantum State: A Multi-Purpose Model for Quantum Many-Body Problems}
% \end{center} 

% \setcounter{page}{1}
\setcounter{figure}{0}
\setcounter{equation}{0} 
\makeatletter
\renewcommand\thefigure{\thesection\arabic{figure}}  
\setcounter{secnumdepth}{3}
\makeatother

% In this supplemental material, we provide all the technical details necessary for reproducing the results in our paper, and show some additional numerical results on the Heisenberg XYZ model \cite{schollwock2008quantum}. 

\section{Transformer implementation details}
\label{sec:implementation}
As illustrated in Fig.~\ref{fig:transformer}, we adopt the standard encoder-only transformer structure \cite{vaswani2017attention}. The discrete spin variables $s_i$ are first one-hot encoded \cite{zheng2018feature}, and the parameters $J_j$ are represented with a scaled one-hot vector. To input interaction strengths and external fields, the scale is the value of the interaction itself. To input the system size $n$, we choose the scale to be $\ln n$ , and append another parity dimension to the input vector, indicating whether $n$ is even or odd. 

Since the input does not entirely consist of one-hot vectors, the embedding layer performs a linear transformation, mapping the input vectors into a $d_e$ dimensional embedding space. 

We use a mixed-style positional encoding. The spin variables $s_i$ have a well-defined position, and we use the $D$-dimensional sinusoidal positional encoding \cite{vaswani2017attention, wang2019translating} on them, where $D$ is the spatial dimension of the physical system. This ensures that the neural network can correctly generalize to larger system sizes it has never been trained on before. On the other hand, the parameters $J_j$ do not have a position, and we use a learnable positional encoding \cite{dosovitskiy2020image} instead. 

After embedding and positional encoding, we pass the embedded inputs through $N$ identical transformer encoder layers, with structures defined in \cite{vaswani2017attention}. The feed-forward sublayer consists of two linear layers, with the hidden dimension in the middle also being $d_e$. We use multi-head self-attention \cite{vaswani2017attention} with 8 heads for the larger model, and 2 heads for the smaller model. ReLU activation \cite{nair2010rectified} is used throughout the neural network. 

After $N$ transformer encoder layers, we use two output heads to model the amplitude and phase of the target wave function. The amplitude head is a linear layer followed by a softmax activation \cite{bridle1990probabilistic}, and the phase head is a linear layer followed by a softsign activation, which is defined in \cite{hibat2020recurrent} and computes the function $(-\infty < x< +\infty)$
\begin{equation}
    \mathrm{softsign}(x)=\frac{x}{1+|x|}.
\end{equation}
We scale the softsign output by $\pi$, to output a phase in the range of $(-\pi, \pi)$. 

The TQS mentioned in the main text has $N=8$ transformer encoder layers with embedding size $d_e=32$, and the number of parameters is about $7.7\times 10^4$. The smaller model in Appendix~\ref{sec:benchmark} has $N=2$ transformer encoder layers with embedding size $d_e=16$, resulting in $5.2\times 10^3$ parameters. The implementation of TQS is carried out using the PyTorch library \cite{paszke2019pytorch}. 

\section{Variational optimization of the ground state energy}
\label{sec:optimization}
TQS is trained by minimizing the ground state energies of a family of Hamiltonians, $\{\hat{H}(\mathbf{J})\}$. 

For a single Hamiltonian $\hat{H}$, the energy derivative reads \cite{carleo2017solving}:

\begin{equation}
    \frac{\partial E}{\partial \theta_k} = 2\mathrm{Re}\bigg(\Big\langle E_{\mathrm{loc}}(\mathbf{s})\frac{\partial \log\psi(\mathbf{s})^*}{\partial \theta_k}\Big\rangle_{P(\mathbf{s})}\bigg)
\end{equation}
where $\langle \cdot \rangle_{P(\mathbf{s})}$ denotes expectation over the distribution $P(\mathbf{s})$, and
\begin{equation}
    E_{\mathrm{loc}}(\mathbf{s}) = \sum_{\mathbf{s}'}\hat{H}(\mathbf{s}, \mathbf{s}')\frac{\psi(\mathbf{s}')}{\psi(\mathbf{s})}
\end{equation}
is the local energy estimator. 

Since the autoregressive wave function is explicitly normalized, it is shown in \cite{hibat2020recurrent} that the variance of the gradient can be reduced by subtracting a baseline energy, 
\begin{equation}
    \frac{\partial E}{\partial \theta_k} = 2\mathrm{Re}\bigg(\Big\langle \big(E_{\mathrm{loc}}(\mathbf{s})-\langle E_{\mathrm{loc}}(\mathbf{s}')\rangle_{P(\mathbf{s}')}\big)\frac{\partial \log\psi(\mathbf{s})^*}{\partial \theta_k}\Big\rangle_{P(\mathbf{s})}\bigg)\label{eq:derivative}
\end{equation}
without introducing bias. This follows from 
\begin{equation}
\begin{aligned}
    &\mathrm{Re}\Big\langle \langle E_{\mathrm{loc}}(\mathbf{s}')\rangle_{P(\mathbf{s}')}\frac{\partial \log\psi(\mathbf{s})^*}{\partial \theta_k}\Big\rangle_{P(\mathbf{s})}\\
    =& \langle E_{\mathrm{loc}}(\mathbf{s}')\rangle_{P(\mathbf{s}')} \sum_{\mathbf{s}}P(\mathbf{s})\frac{1}{2}\frac{1}{P(\mathbf{s})}\frac{\partial P(\mathbf{s})}{\partial \theta_k}\\
    =& \frac{E}{2}\frac{\partial}{\partial \theta_k}\sum_{\mathbf{s}} P(\mathbf{s}) = \frac{E}{2}\frac{\partial}{\partial \theta_k} 1 = 0.
\end{aligned}
\end{equation}

In our problem setting, we have a family of Hamiltonians $\hat{H}(\mathbf{J})$ parameterized by $\mathbf{J}$, with ground state energies $E_g(\mathbf{J})$. Without loss of generality, we suppose all $E_g<0$; otherwise we can simply shift the energy levels by adding a constant. Then, we define the super-Hamiltonian, 
\begin{equation}
    \hat{\mathcal{H}} = \bigoplus_{\mathbf{J}} \frac{\hat{H}(\mathbf{J})}{|E_g(\mathbf{J})|}, 
\end{equation}
to be the direct sum of all (possibly infinite) Hamiltonians $H(\mathbf{J})$, weighted by their ground state energies, $\frac{1}{|E_g(\mathbf{J})|}$. Note that $\hat{\mathcal{H}}$ is block diagonal, with no interaction across different $\mathbf{J}$. One can easily show that, the ground state of $\hat{\mathcal{H}}$ is the direct sum of all ground states,  $|\Psi\rangle=\bigoplus_{\mathbf{J}}|\psi(\mathbf{J})\rangle$,
\begin{equation}
\begin{aligned}
    \hat{\mathcal{H}}|\Psi\rangle =& \Bigg(\bigoplus_{\mathbf{J}}\frac{\hat{H}(\mathbf{J})}{|E_g(\mathbf{J})|}\Bigg)\Bigg(\bigoplus_{\mathbf{J}}|\psi(\mathbf{J})\rangle\Bigg)\\
    =& \bigoplus_{\mathbf{J}}\frac{\hat{H}(\mathbf{J})|\psi(\mathbf{J})\rangle}{|E_g(\mathbf{J})|}\\
    =& -\bigoplus_{\mathbf{J}}|\psi(\mathbf{J})\rangle = -|\Psi\rangle
\end{aligned}
\end{equation}
with eigenvalue $-1$. Therefore, we can follow the standard procedure and minimize 
\begin{equation}
    \langle \Psi|\hat{\mathcal{H}}|\Psi\rangle =  \sum_{\mathbf{J}}\frac{\langle\psi(\mathbf{J})|\hat{H}(\mathbf{J})|\psi(\mathbf{J})\rangle}{|E_g(\mathbf{J})|}.
\end{equation}

We don't have access to the exact ground state energies $E_g(\mathbf{J})$, so we instead approximate them with variationally approximated ground state energies, $\tilde{E}_g(\mathbf{J})=\langle E_{\mathrm{loc}}(\mathbf{s}, \mathbf{J})\rangle$, which become increasingly more accurate as optimization goes on. 

In practice, at each optimization iteration, we sample a random $\mathbf{J}$ according to $P(\mathbf{J})$, and compute the energy derivative Eq.~\eqref{eq:derivative}, scaled by $\frac{1}{|\tilde{E}_g(\mathbf{J})|}$.  We set an upper limit of 5 to the scaling factor, to avoid divergences when $\tilde{E}_g(\mathbf{J})\to 0$ during optimization. 

The entire training procedure is carried out using the Adam optimizer \cite{kingma2014adam}, with $\beta_1=0.9$ and $\beta_2=0.98$. We varied the learning rate during training according to the formula, 
\begin{equation}
    \mathrm{lr}(i_{\mathrm{step}}) = 5 d_e^{-0.5} \min(i_{\mathrm{step}}^{-0.75}, i_{\mathrm{step}}i_{\mathrm{warmup}}^{-1.75} )
\end{equation}
where $d_e$ is the embedding size of the model, $i_{\mathrm{step}}$ is the current number of training steps, and $i_{\mathrm{warmup}}$ is the number of warm up steps. We used $i_{\mathrm{warmup}}=4000$. This corresponds to linearly increasing the learning rate during the first 4000 iterations, and polynomially decreasing it during the rest of the training. This learning rate schedule is inspired from \cite{vaswani2017attention}. During fine-tuning, we use a different learning rate schedule: 
\begin{equation}
    \mathrm{lr}(i_{\mathrm{step}}) = 5 d_e^{-0.5} (i_{\mathrm{step}}+10^5)^{-0.75}.
\end{equation}

\section{Sampling algorithm}
\label{sec:sampling}
The autoregressive structure of TQS already makes sampling efficient, and the efficiency is further improved by adopting the sampling algorithm in \cite{barrett2022autoregressive}, which only samples unique configuration strings. 

During sampling, we first fix a large batch size, $N_{\mathrm{batch}}$, and autoregressively sample the spins to form partial strings, $\mathbf{s}^k=s_1 s_2 \cdots s_i$, with associated number of occurrences, $n_k$. At the $(i+1)$-th sampling step, $s_{i+1}$ is sampled from the conditional distribution $P(s_{i+1}|s_1, \cdots, s_{i}, \mathbf{J})$, resulting in $n_{k0}$ occurrences of $s_{i+1}=0$ and $n_{k1}$ occurrences of $s_{i+1}=1$, with $n_{k0}+n_{k1}=n_k$. After this step, we obtain two unique partial strings, $\mathbf{s}^{k0}=s_1 s_2 \cdots s_i 0$ and $\mathbf{s}^{k1}=s_1 s_2 \cdots s_i 1$, with occurrences $n_{k0}$ and $n_{k1}$, respectively. This procedure starts from an empty set and is repeated until the number of unique strings reaches a maximum, $N_{\mathrm{unique}}$, after which no new partial string branches are generated, and the remaining spins are sampled in the regular way. 

The complexity of this sampling algorithm is approximately proportional to $N_{\mathrm{unique}}$ and does not depend on $N_{\mathrm{batch}}$. Therefore, we can choose extremely large batch sizes to greatly improve on the accuracy of estimated expectation values, with negligible increase in computation time. For all the experiments mentioned in this paper, we choose $N_{\mathrm{batch}}=10^6$,  $N_{\mathrm{unique}}=10^2$ during training, and $N_{\mathrm{unique}}=10^3$ during evaluations. 

\section{Implementing symmetries}
\label{sec:symmetry}
The transformer architecture itself does not observe any symmetry, but most Hamiltonians do. To impose symmetries without spoiling the autoregressive structure, we follow the approaches in previous works \cite{choo2018symmetries, hibat2020recurrent, sharir2020deep} and explicitly symmetrize the wave function in a similar way. 

Suppose $\hat{\mathcal{T}}$ is a discrete symmetry of $\hat{H}$, with $\hat{\mathcal{T}}^m=\mathds{1}$ ($m\in \mathbb{N}$). By definition, we have $[\hat{H}, \hat{\mathcal{T}}]=0$, and one can simultaneously diagonalize both operators within the same eigenbasis. Under this basis, the ground state $|\psi\rangle$ is also an eigenstate of $\hat{\mathcal{T}}$, 
\begin{equation}
    \hat{\mathcal{T}}|\psi\rangle = \omega_{\hat{\mathcal{T}}}|\psi\rangle \label{eq:T}
\end{equation}
where $\omega_{\hat{\mathcal{T}}}=e^{2\pi i k/m}$, $k\in \mathbb{N}$. Expanding Eq.~\eqref{eq:T} in the computational basis, we get
\begin{equation}
    \psi(\hat{\mathcal{T}}^{-1}\mathbf{s}) = \omega_{\hat{\mathcal{T}}}\psi(\mathbf{s}).\label{eq:symmetry}
\end{equation}

In terms of amplitude and phase, Eq.~\eqref{eq:symmetry} becomes
\begin{equation}
\begin{aligned}
    A(\hat{\mathcal{T}}\mathbf{s}) &= A(\mathbf{s}),\\
    \phi(\hat{\mathcal{T}}\mathbf{s}) &= \phi(\mathbf{s})-\frac{2\pi k}{m}.
\end{aligned}\label{eq:symm_amp_phase}
\end{equation}

The output wave function from TQS clearly does not satisfy Eq.~\eqref{eq:symm_amp_phase}. To explicitly enforce the symmetry $\hat{\mathcal{T}}$, we define
\begin{equation}
\begin{aligned}
    \tilde{P}(\mathbf{s}) &= \frac{1}{m}\sum_{n=0}^{m-1}P(\hat{\mathcal{T}}^n \mathbf{s})\\
    \tilde{\phi}(\mathbf{s}_0) &= \mathrm{Arg}\Bigg(\sum_{n=0}^{m-1}\psi(\hat{\mathcal{T}}^n\mathbf{s}_0)\Bigg)\\
    \tilde{\phi}(\hat{\mathcal{T}}^n\mathbf{s}_0) &= \tilde{\phi}(\mathbf{s}_0) - \frac{2\pi k n}{m}
\end{aligned}\label{eq:symmetrized_wave_function}
\end{equation}
where $\psi(\mathbf{s})=\sqrt{P(\mathbf{s})}e^{i\phi(\mathbf{s})}$, $P$, $\phi$ are outputs from the TQS, and $\tilde{P}$, $\tilde{\phi}$ are symmetrized probability and phase, respectively. $\mathbf{s}_0$ is an arbitrary initial configuration in each symmetry sector, predefined so that the phases within the symmetry sector can be assigned consistently. We choose $\mathbf{s}_0$ to be the configuration with the smallest decimal value, converted from its binary bitstring, within each symmetry sector. 

Sampling from the symmetrized wave function has almost no additional computational cost. We follow the same procedure detailed in the previous section, and apply a random symmetry operation $\hat{\mathcal{T}}^n$ to the sampled configuration $\mathbf{s}$ in the end \cite{sharir2020deep, hibat2020recurrent}. However, to compute the exact value of ${\psi}(\mathbf{s})$, one needs to evaluate all configurations within the symmetry sector and explicitly calculate Eq.~\eqref{eq:symmetrized_wave_function}, which is $m$ times more expensive.

Another symmetry worth mentioning is the $U(1)$ symmetry of the Heisenberg model, which leads to zero magnetization. This symmetry is particularly easy to implement, and we follow the same method developed in \cite{hibat2020recurrent}, by setting the probability of a partial string to 0 whenever the number of up spins or down spins exceeds half of the system size.

Note that, while the Hamiltonian $\hat{H}$ may satisfy several symmetries $\hat{\mathcal{T}}_1, \hat{\mathcal{T}}_2, \cdots$, it is possible that $[\hat{\mathcal{T}}_1, \hat{\mathcal{T}}_2]\neq 0$, making it impossible to diagonalize all symmetries at the same time. However, this does not pose a problem for us. Although $\hat{\mathcal{T}}_1$ and $\hat{\mathcal{T}}_2$ do not commute in general, they do commute in certain symmetry sectors (for example, $\omega_{\hat{\mathcal{T}}_1}=\omega_{\hat{\mathcal{T}}_2}=1$). To implement symmetries, we need to know $\omega_{\hat{\mathcal{T}}}$ as a prior knowledge, and this information then helps us determine all compatible symmetries. 

As another remark, in our implementation of TQS, we only enforced the symmetries $\hat{\mathcal{T}}$ with $\omega_{\hat{\mathcal{T}}}=1$. We noticed that, any symmetry with $\omega_{\hat{\mathcal{T}}}\neq 1$ would impose a non-trivial phase structure to the wave function, which is somewhat arbitrary and could significantly slow down the training. 

\section{Predicting parameters}
\label{sec:param}
With the learned distribution $P(\mathbf{s}, \mathbf{J})$, we can predict the parameters $\mathbf{J}$ from a batch of measurements $\{\mathbf{s}_i\}$. As illustrated in the main text, this is achieved through maximizing the log-likelihood functional Eq.~\eqref{eq:log_likelihood}. 

We carried out the maximization using the Nelder-Mead method \cite{nelder1965simplex}, a heuristic searching algorithm based on a moving simplex, implemented in the SciPy library \cite{2020SciPy-NMeth}, with a tolerance of $10^{-9}$. 

An alternative method to predict the parameters would be supervised  fine-tuning, which adds a parameter prediction head as an additional output of TQS. This would have the advantage of reducing the computational cost to one forward pass, at the expense of fine-tuning cost. We leave this as a future work. 

Note that we didn't use any phase information during the prediction. To make use of the phase structure, one needs to perform measurements in different bases, and compute a generalized likelihood function that takes all bases into account. For this, we refer the readers to \cite{torlai2018neural}. Alternatively, it is possible to use informationally-complete positive operator-valued measurements (IC-POVM) to encode the complete information of a quantum state, which is developed in \cite{carrasquilla2019reconstructing}. We can adapt the TQS structure to be compatible with IC-POVM, which we also leave as a future work. \\

\section{DMRG calculations}
\label{sec:DMRG}
For the 1D transverse field Ising model in the main text and the 1D XYZ model in Appendix~\ref{sec:XYZ}, we use density matrix renormalization group (DMRG) as a benchmark to evaluate the performance of our algorithm. DMRG can be extremely accurate for 1D systems, yet performs rather poorly in 2 or more dimensions \cite{schollwock2011density}.

We used the TeNPy library \cite{tenpy} to perform DMRG calculations. For all DMRG results mentioned in the paper, we use a maximum bond dimension of 100, and terminate when the energy tolerance $10^{-10}$ is achieved. 

\section{Additional numerical results}
\subsection{Performance benchmarking}
\label{sec:benchmark}
In this section, we compare the accuracy and training cost of TQS with restricted Boltzmann machine (RBM) \cite{carleo2017solving}, another widely adopted framework for neural network quantum states. 

To ensure a fair comparison, we trained a smaller TQS with $5.2\times 10^3$ parameters (embedding size $d_e=16$, two transformer encoder blocks), to compare with an RBM with approximately the same number of parameters (hidden-to-variable ratio $\alpha=3$). The TQS is trained using the setting described in the previous sections, with $N_{\mathrm{unique}}=2000$, while the RBM is trained using stochastic reconfiguration (SR) \cite{sorella2007weak, carleo2017solving}, contrastive divergence with 10 sampling steps (CD-10) \cite{carreira2005contrastive} and batch size 24800. Under this setting, the computational cost for each training iteration is approximately the same. To model continuous physical parameters in RBMs, we use continuous visible neurons normalized to $[-2, 2]$, together with regular binary neurons with values $\pm 1$ for the spin variables. 

\begin{figure}[htbp]
	\centering
	\includegraphics[width = 0.45\textwidth]{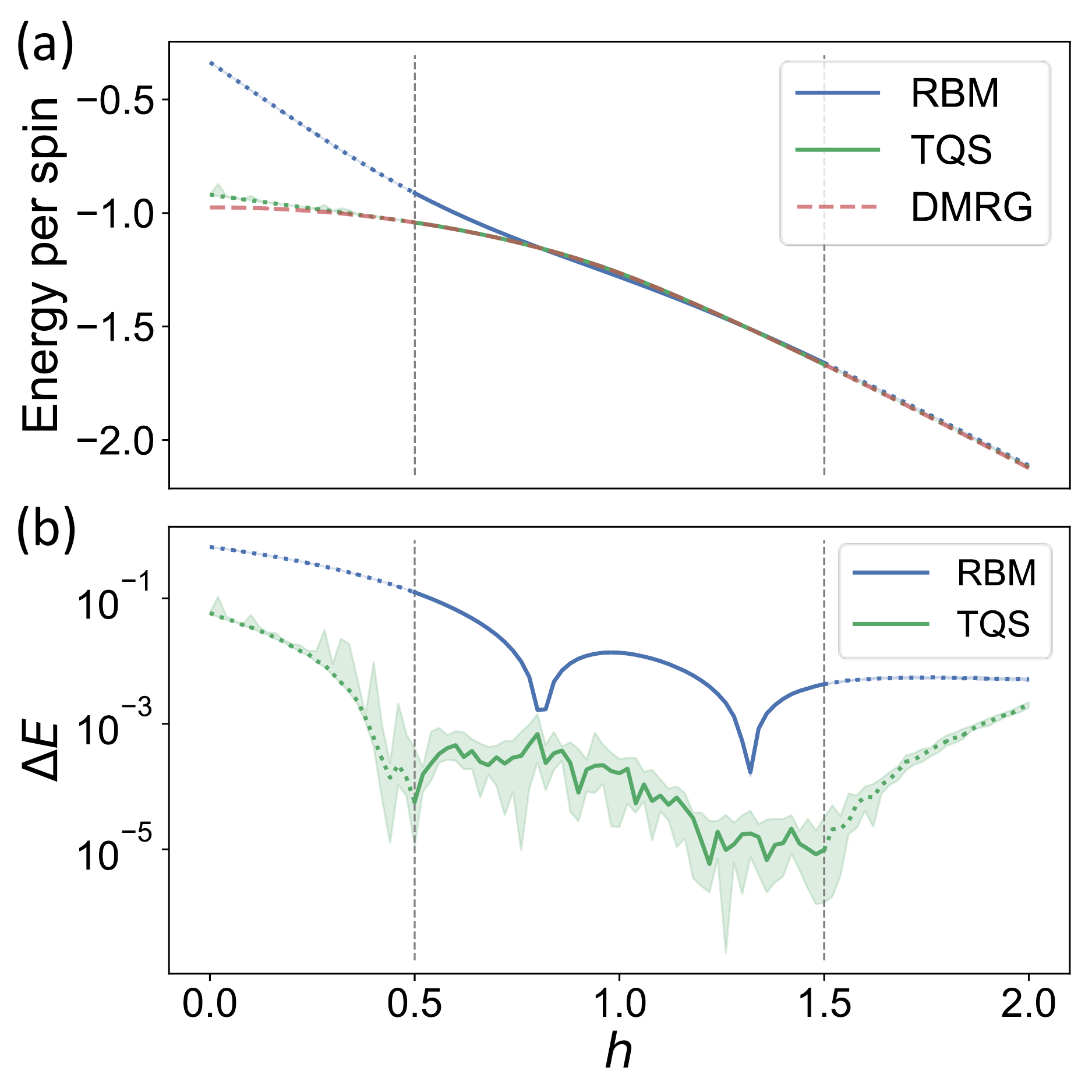}
	\caption{Comparison of TQS and RBM on the ground state of the TFI model, with a variable transverse field $h$. (a) Energy per spin and (b) Relative error of the ground state energy, $\Delta E=|(E-E_{\mathrm{ground}})/E_{\mathrm{ground}}|$. Both models are trained in the range $h\in[0.5, 1.5]$.}
	\label{fig:E_mid}
\end{figure}

\begin{figure}[htbp]
	\centering
	\includegraphics[width = 0.45\textwidth]{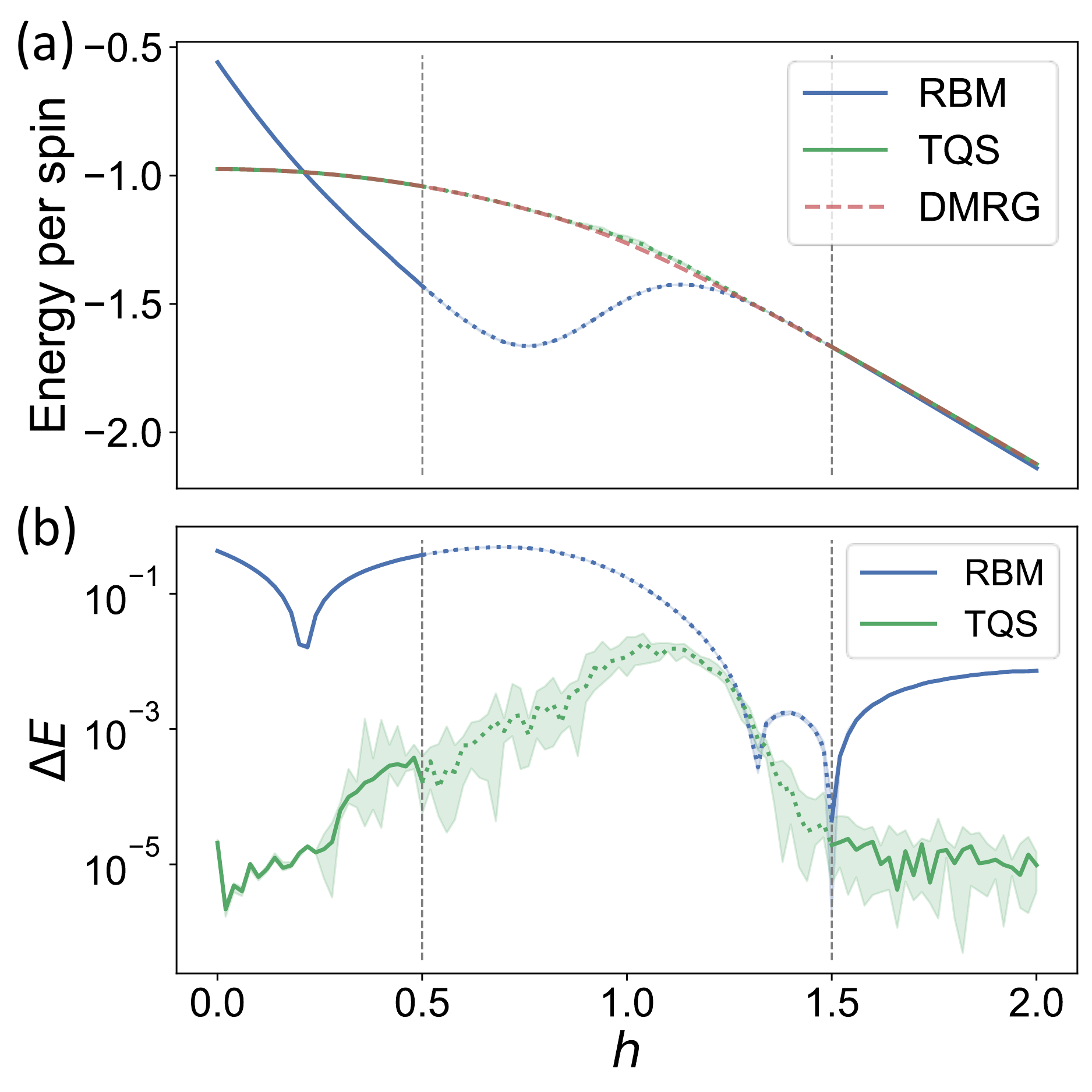}
	\caption{Comparison of TQS and RBM on the ground state of the TFI model, with a variable transverse field $h$. (a) Energy per spin and (b) Relative error of the ground state energy. Both models are trained in the range $h\in[0, 0.5]\cup[1.5, 2]$. }
	\label{fig:E_phase_transition}
\end{figure}

At this point, we try to reproduce the experiment described in the main text using RBMs. We focus on the transverse field Ising (TFI) model, with the transverse field $h$ as an additional input to the neural network. The RBM is trained for $10^5$ iterations, and the learning rate decreases according to the formula
\begin{equation}
    \mathrm{lr}(i_{\mathrm{step}})=\mathrm{lr}_{\max}i_{\mathrm{step}}^{-0.5}
\end{equation}
where $i_{\mathrm{step}}$ is the current number of training steps, and $\mathrm{lr}_{\max}=0.02$. TQS is also trained for $10^5$ iterations, and the results are shown in Figs.~\ref{fig:E_mid}, \ref{fig:E_phase_transition}. 

In Fig.~\ref{fig:E_mid}, the training range is $h\in[0.5, 1.5]$, and the RBM learned an energy curve that is almost linear in $h$. And in Fig.~\ref{fig:E_phase_transition} the training range is $h\in[0, 0.5]\cup[1.5, 2]$, but the RBM only learned the properties in the $[1.5, 2]$ range. In comparison, TQS did an almost perfect job in both cases. 

This result is expected, since TQS is designed for flexibility and is able to learn different quantum states at the same time, even with a tiny model size. On the other hand, RBM can accurately represent a single quantum state, but it is much less flexible when it comes to a family of quantum states. 

\begin{figure}[htbp]
	\centering
	\includegraphics[width = 0.45\textwidth]{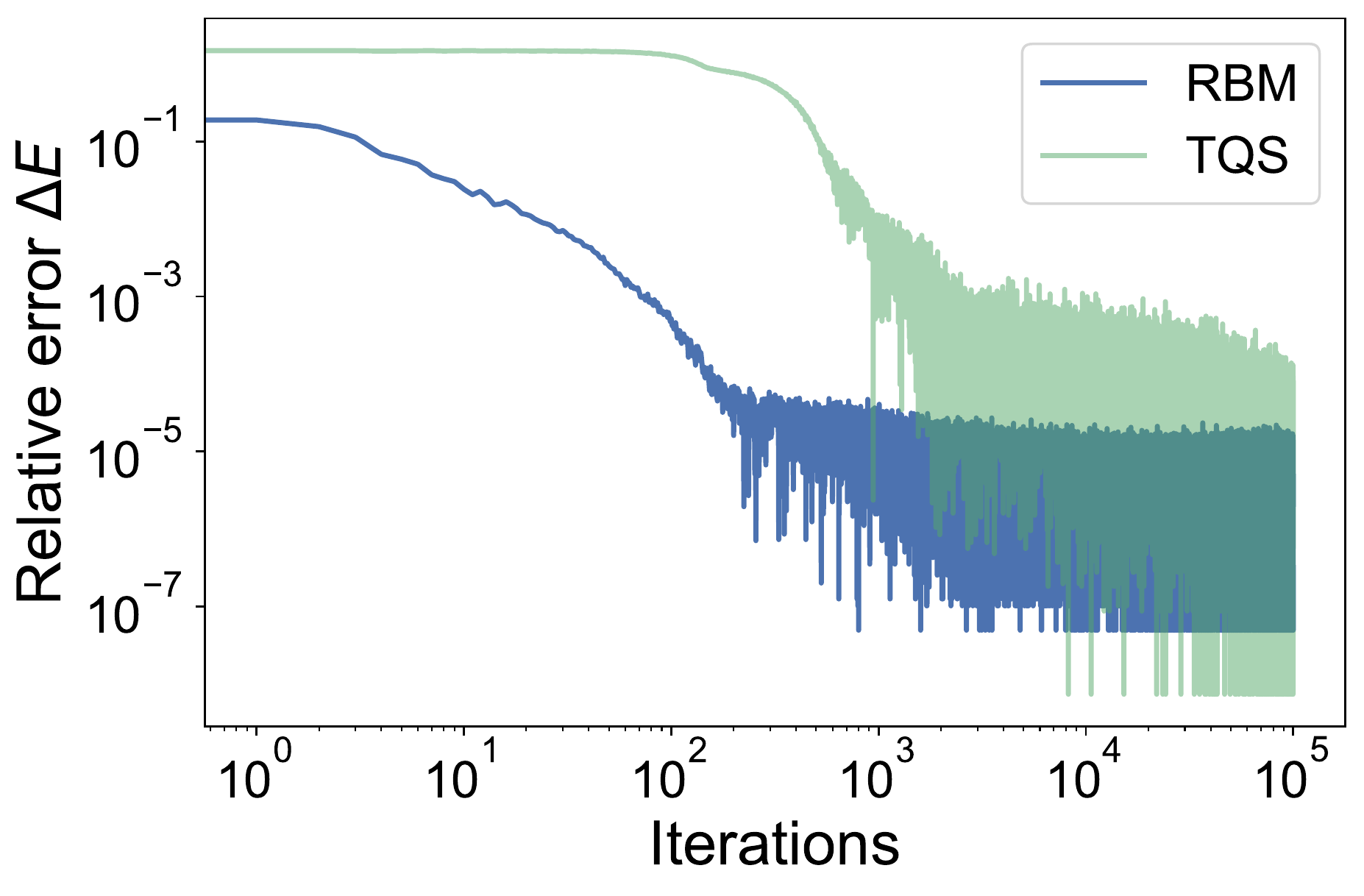}
	\caption{The training curve of TQS and RBM on the ground state of the TFI model with transverse field $h=1$. On a single data point, RBM converges faster than TQS. }
	\label{fig:energy_curve}
\end{figure}

As another test, we train both models at a single data point $h=1$ for $10^5$ iterations, and the result is shown in Fig.~\ref{fig:energy_curve}. RBM works very well in this case, converging in about $10^3$ iterations, and does not improve much afterwards. On the other hand, TQS converges much slower, but continues to see improvements up to $10^5$ iterations. 

Again, this result is expected. With a simple structure, RBM can be easily trained using the SR algorithm, leading to a fast convergence. However, TQS has a much more sophisticated structure, and needs a warm-up period in the learning rate schedule for a smoother convergence. This makes the training slower, but with a potentially higher final accuracy. 

\subsection{Finite-size scaling}
\label{sec:finite_size}
The idea of finite-size scaling \cite{fisher1972scaling} can help us understand divergent behaviors in the thermodynamic limit using only numerical results in finite systems. Assume we have some physical quantity $\Omega$ that diverges in the thermodynamic limit at a critical value $h_c$, 
\begin{equation}
    \Omega(h)\sim |\Delta h|^{-\omega},
\end{equation}
where $\Delta h=(h-h_c)/h_c\to 0$. The correlation length, $\xi(h)\sim|\Delta h|^{-\nu}$, also diverges with critical exponent $\nu$. Therefore, $\Omega$ correlates with $\xi$ as
\begin{equation}
    \Omega\sim\xi^{\omega/\nu}. \label{eq:inf_system}
\end{equation}

For a finite system of linear size $N$, the behavior of $\Omega(h, N)$ deviates according to the ratio $\xi/N$. When $\xi\ll N$, finite-size effects are negligible, and Eq.~\eqref{eq:inf_system} is preserved. However, since the correlation length cannot exceed the system size in finite systems, if $\xi\gg N$, $\Omega$ has to scale with $N$ instead. This leads to the finite-size scaling ansatz
\begin{equation}
    \Omega(h, N)\sim \xi^{\omega/\nu} f(N/\xi), \label{eq:finite}
\end{equation}
where $f(x)$ is a scaling function that satisfies
\begin{equation}
f(x)\sim\begin{cases}
\mathrm{const}, & x\to\infty, \\
x^{\omega/\nu}, & x\to 0.
\end{cases}
\end{equation}

By defining $g(x)=x^{-\omega}f(x^\nu)$, we can rewrite Eq.~\eqref{eq:finite} as
\begin{equation}
    \Omega(h, N)\sim N^{\omega/\nu} g(N^{1/\nu}|\Delta h|)
\end{equation}
Therefore, at the critical point $h_c$, $\Omega$ scales as
\begin{equation}
    \Omega(h_c, N)\sim N^{\omega / \nu}.\label{eq:system_size_scaling}
\end{equation}

Determining $h_c$ is another task on its own. A common method is to compute the Binder cumulant \cite{binder1981finite},
\begin{equation}
    U_N = 1 - \frac{\langle m^4_z\rangle_N}{3\langle m^2_z\rangle^2_N},
\end{equation}
which is invariant with system size $N$ at the critical point \cite{binder1981finite}. Therefore, the crossing point of $U_N-h$ curves for different $N$ gives the critical point $h_c$. 

In Fig.~\ref{fig:finite_size}(a) in the main text, we used the Binder cumulant to show that TQS can correctly identify the TFI phase transition at $h=1$. And in Fig.~\ref{fig:finite_size}(b), we computed the ratio $\beta/\nu$ by fitting the scaling of magnetization $m_z$ to Eq.~\eqref{eq:system_size_scaling}. On a side note, since TQS is explicitly symmetrized to have $\langle m_z\rangle=0$, we followed the method in \cite{pang2019critical} and used the mean-square-root magnetization $\sqrt{\langle m_z^2\rangle}$ instead. 

\begin{figure}[htbp]
	\centering
	\includegraphics[width = 0.45\textwidth]{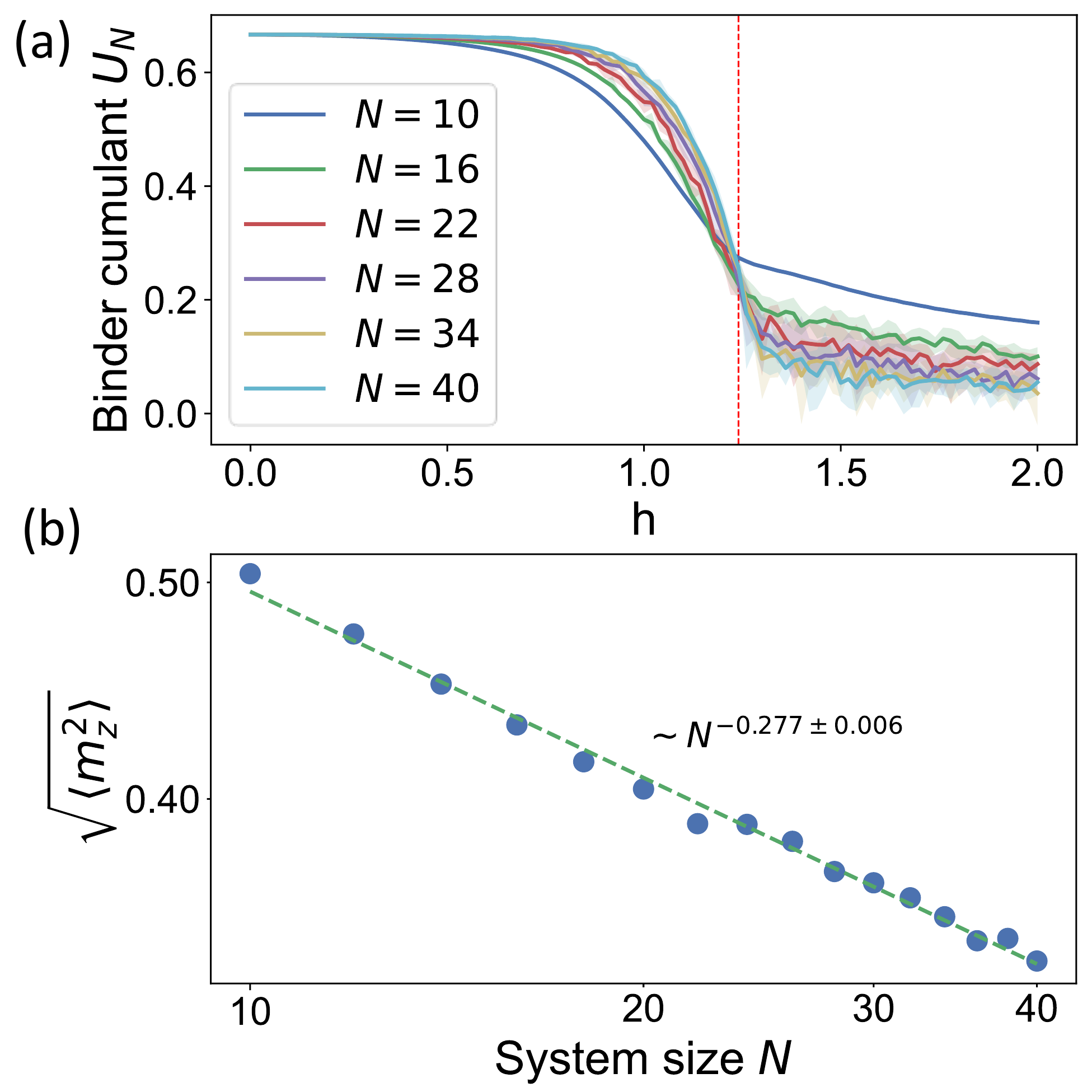}
	\caption{Finite-size scaling calculations on the TFI model, using the TQS trained in $h\in[0, 0.5]\cup[1.5, 2]$. (a) Binder cumulant \cite{binder1981finite}, $U_N=1-\frac{\langle m_z^4\rangle_N}{3\langle m_z^2\rangle_N^2}$, plotted for various system sizes $N$. The curves cross at $h=1.24$. (b) Finite-size scaling of the mean-square-root magnetization at the critical point $h=1.24$. A linear fit on the log-log scale gives $\beta/\nu=0.277\pm0.006$. } 
	\label{fig:finite_size_edge}
\end{figure}

The results in Fig.~\ref{fig:finite_size} are obtained using a TQS model trained in $h\in[0.5, 1.5]$ near the phase boundary. What if the TQS has never been trained on any data near the critical point? To test this, we performed the same analysis using TQS trained in $h\in[0, 0.5]\cup[1.5, 2]$, either deep in the paramagnetic or ferromagnetic regime. The results are shown in Fig.~\ref{fig:finite_size_edge}. This time, TQS failed to find the correct critical point $h_c=1$. However, quite surprisingly, TQS managed to find a plausible interpolation between the two phases, with a new critical point $h=1.24$, and critical exponents $\beta/\nu=0.277\pm 0.006$. 

Of course, this interpolated phase transition is not physical, and the computed critical exponents seem to suggest that this fictitious system has a fractal dimension between 1 and 2. It would be an interesting future work to look into the neural network and analyze what actually happened here. 

\subsection{Heisenberg XYZ model}
\label{sec:XYZ}
In this section, we further benchmark the performance of TQS with additional numerical experiments. We focus on the 1D Heisenberg XYZ model in a longitudinal field \cite{schollwock2008quantum}, whose Hamiltonian is given by

\begin{equation}
\begin{aligned}
    \hat{H} = J \sum_{i=1}^{n-1} \Big[ & (1+\gamma)\sigma^x_i \sigma^x_{i+1} + (1-\gamma)\sigma^y_i \sigma^y_{i+1} \\
    +& \Delta \sigma^z_i \sigma^z_{i+1}  \Big]
    + h \sum_{i=1}^n \sigma^z_i    
\end{aligned}
\end{equation}

\begin{figure}[H]
	\centering
	\includegraphics[width = 0.45\textwidth]{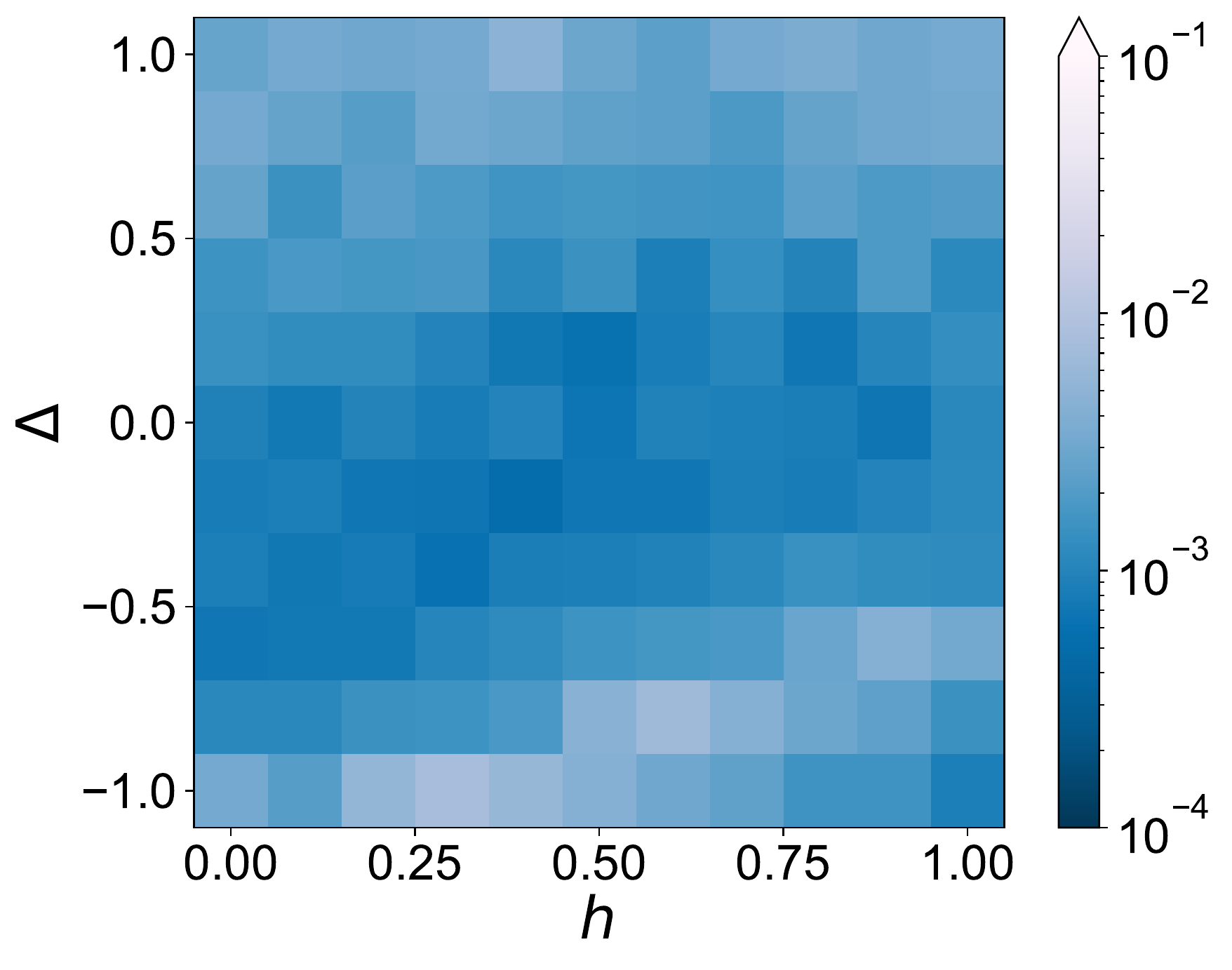}
	\caption{The relative error of the ground state energy, $|(E-E_{\mathrm{ground}})/E_{\mathrm{ground}}|$, plotted against the external field $h$ and longitudinal interaction strength $\Delta$, with $n=40$. In this figure, $h$ and $\Delta$ are in the pre-training range. }
	\label{fig:XYZ_pt}
\end{figure}

\begin{figure}[H]
	\centering
	\includegraphics[width = 0.45\textwidth]{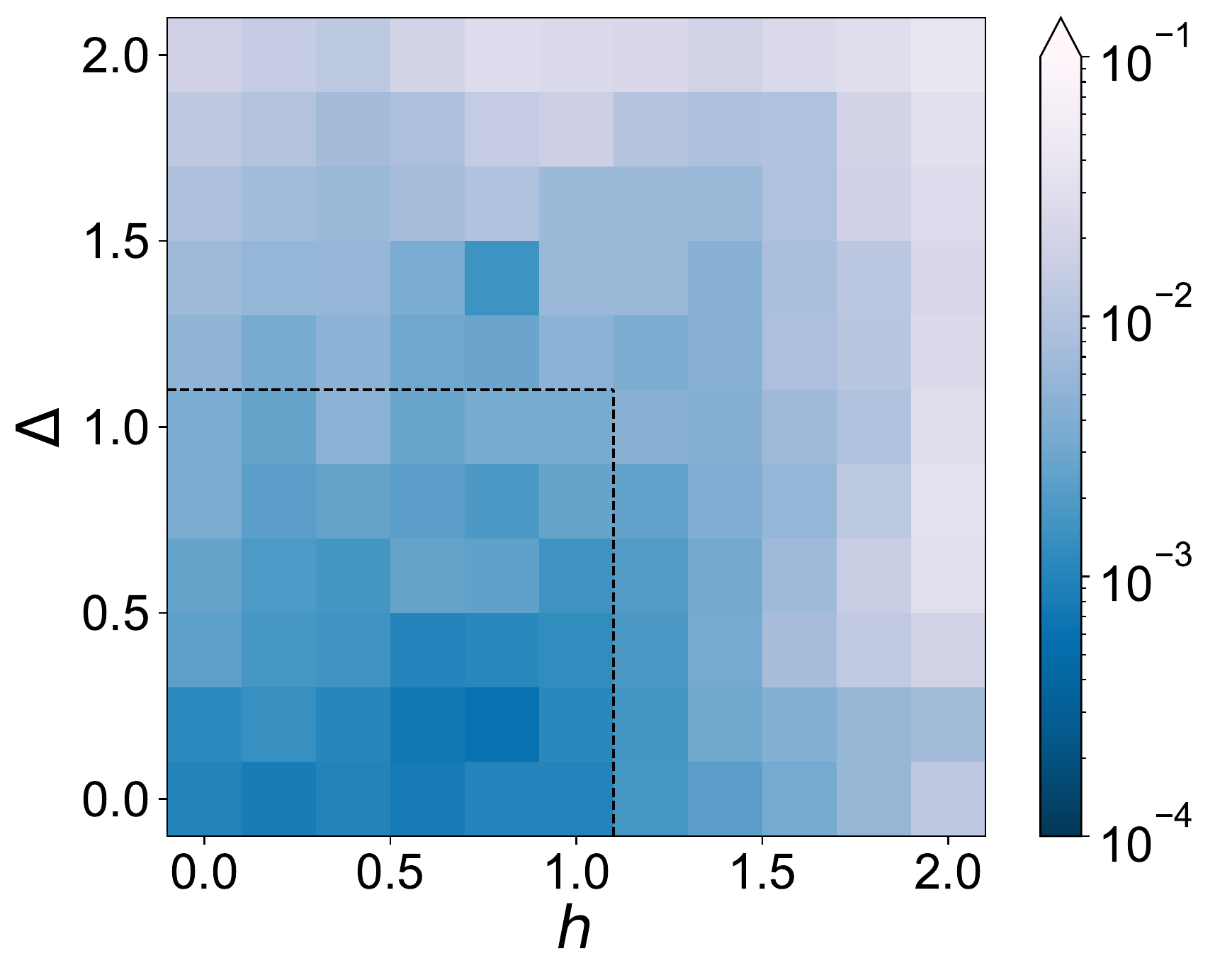}
	\caption{The relative error of the ground state energy, $|(E-E_{\mathrm{ground}})/E_{\mathrm{ground}}|$, in an extended parameter range. The bottom left corner is part of the pre-training range, separated with black dashed lines for visual clarity.}
	\label{fig:XYZ_extended}
\end{figure}

We fix $J=1, \gamma=0.2$, and consider the parameter range $h\in[0, 1]$, $\Delta\in[-1, 1]$. The system size $n$ can take any even integer value with equal probability in the range of $[10, 40]$. The TQS has the same structure as the one described in the main text, with 8 layers and embedding size 32. We trained the TQS for $10^5$ iterations without implementing any symmetry, and the relative errors of the ground state energy, $|(E-E_{\mathrm{ground}})/E_{\mathrm{ground}}|$, for system size $n=40$, are plotted in Figs.~\ref{fig:XYZ_pt}, \ref{fig:XYZ_extended}. 

Fig.~\ref{fig:XYZ_pt} shows the results in the pre-trained range, $h\in[0, 1], \Delta\in[-1, 1]$, and the accuracy is at the order of $10^{-3}$. In Fig.~\ref{fig:XYZ_extended}, we extended the parameter range to $h\in[0, 2], \Delta\in[0, 2]$, with pre-trained and extended parameter ranges separated by black dashed lines. TQS can still reasonably infer the ground state properties outside of the pre-trained range, but its accuracy gradually decreases as we move further away. 

\begin{figure}[htbp]
	\centering
	\includegraphics[width = 0.45\textwidth]{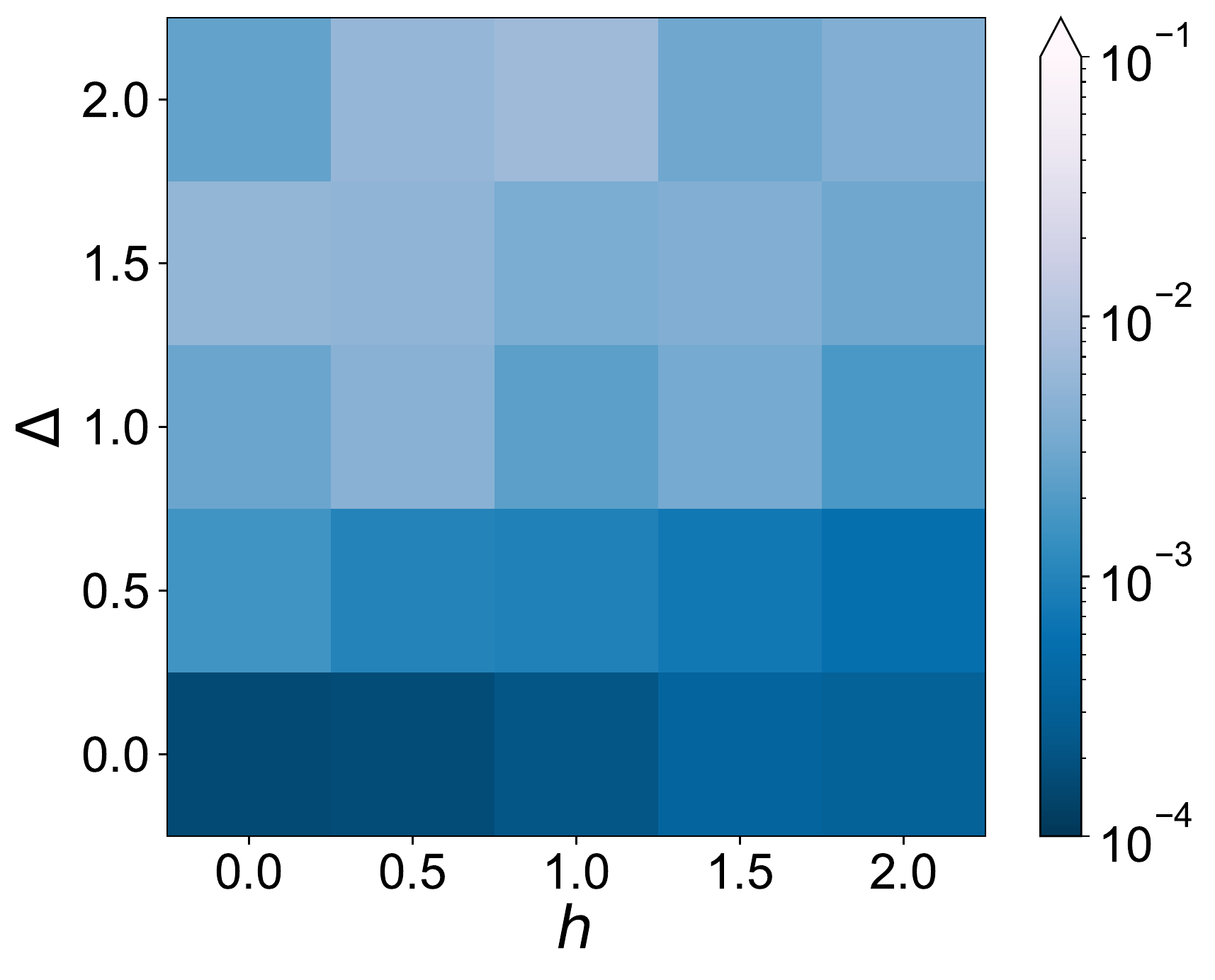}
	\caption{The relative error of the ground state energy, $|(E-E_{\mathrm{ground}})/E_{\mathrm{ground}}|$, after fine-tuning the TQS on specific parameter points for $2\times 10^3$ iterations. }
	\label{fig:XYZ_ft}
\end{figure}

To improve on the energy estimations, we fine-tune the pre-trained TQS at selected parameters for $2\times 10^3$ iterations. The results are plotted in Fig.~\ref{fig:XYZ_ft}. With fine-tuning, the ground state energy accuracy improved by another order of magnitude, allowing us to more accurately estimate the ground state properties on a much wider parameter range, with minimal computational cost.

\end{document}